\documentclass[iop]{emulateapj}
\def\maxi{{\it MAXI}}
\def\pca{{PCA}}
\def\RXTE{{\it RXTE}}
\def\Chandra{${\it Chandra}$}

\newcommand{\Msun}{\ifmmode {M_{\odot}}\else${M_{\odot}}$\fi}

\usepackage{graphicx}

\newcommand{\lsim }{{\lower0.8ex\hbox{$\buildrel <\over\sim$}}}
\newcommand{\gsim }{{\lower0.8ex\hbox{$\buildrel >\over\sim$}}}

\shorttitle{Ultracompact X-ray Binary Luminosities}
\shortauthors{Cartwright et al.}

\begin{document}
\title{Galactic Ultracompact X-ray Binaries: Empirical Luminosities }
\author{T.~F. Cartwright\altaffilmark{1}, M.~C. Engel, C.~O. Heinke\altaffilmark{2}, G.~R. Sivakoff, J.~J. Berger, J.~C. Gladstone,  N.~Ivanova}
\affil{Physics Dept., 4-183 CCIS, Univ. of Alberta, Edmonton AB, T6G 2E1, Canada}
\altaffiltext{1}{Now at International Space University, 1 rue Jean-Dominique Cassini, 67400 Illkirch-Graffenstaden, France}
\altaffiltext{2}{Ingenuity New Faculty; heinke@ualberta.ca}

\begin{abstract}
Ultracompact X-ray binaries (UCXBs) are thought to have relatively simple binary evolution post-contact, leading to clear predictions of their luminosity function. We test these predictions by studying the long-term behavior of known UCXBs in our Galaxy, principally using data from the \maxi\ All-Sky Survey and the Galactic Bulge Scans with {\it RXTE's} \pca\ instrument.  
Strong luminosity variations are common (and well-documented) among persistent UCXBs, which requires an  explanation other than the disk instability mechanism.  We measure the luminosity function of known UCXBs in the Milky Way, which extends to lower luminosities than some proposed theoretical luminosity functions of UCXBs.  
The difference between field and globular cluster X-ray luminosity functions in other galaxies cannot be explained by an increased fraction of UCXBs in globular clusters.  Instead, our measured luminosity function suggests that UCXBs only make up a small fraction of the X-ray binaries above a few $\times10^{36}$ ergs/s in both old field populations and globular clusters.
\end{abstract}



\maketitle

\section{Introduction}\label{s:intro}

Ultra-compact X-ray binaries (UCXBs) contain a compact accretor star (in all known cases, a neutron star [NS], though black hole UCXBs are possible) and a compact donor star, with an orbital period $P_{\rm orb}<80$ minutes. Such short periods require the donors to be hydrogen-deficient, partially or fully degenerate stars \citep[e.g.][]{Rappaport82,Deloye03}.
UCXB systems are more common in globular clusters (GCs), likely due to their formation there by close dynamical interactions \citep{Verbunt87b,Deutsch00,Ivanova05}. 

UCXB systems can be roughly divided into persistent (over the decades we have been observing them) and  transient systems.  Transient UCXBs spend the majority of the time in a quiescent state with little or no accretion, punctuated by occasional outbursts (when they become quite luminous, and are more easily detected).  The outbursts of known transient UCXBs are rather short, which is expected, given the small size of their accretion disks.

Chandra X-ray observations of elliptical galaxies detect large numbers of X-ray binaries in GCs \citep[e.g.][]{Sarazin01,Angelini01,Kundu02,Minniti04,Jordan04}.  
 The luminosity function of GC X-ray binaries differs from the non-cluster X-ray binary luminosity function, particularly below an X-ray luminosity $L_X=10^{37}$ ergs/s where GCs have fewer X-ray binaries \citep{Fabbiano07,Kim09,Voss09,Zhang11}.  UCXBs have been suggested to dominate bright GC X-ray sources \citep{Bildsten04}, and their high frequency in GCs has been suggested to explain the different X-ray luminosity functions inside vs. outside of clusters \citep{Voss09}.  This suggestion depends on the \citet{Fragos08} model of the UCXB luminosity function, which cuts off below 5$\times10^{36}$ ergs s$^{-1}$ (following current calculations of the He disk instability line), to explain the flat luminosity function of GCs at low $L_X$. 
\citet{Ivanova08} found that 
UCXBs are unlikely to constitute the majority of GC X-ray binaries due to the short lifetimes of persistent UCXBs; instead, main sequence donors are preferred for that role.
 An empirical measurement of the UCXB luminosity function (along with re-consideration of the appropriate $\dot{M}_{\rm crit}$) can uncover the true role of UCXBs among the X-ray binaries in elliptical galaxy GCs, and is a target of this work.

 In this paper, we compile histograms of the luminosities of galactic UCXBs from the most sensitive surveys available, and calculate empirical UCXB luminosity functions.  In \citet{Heinke12a}, hereafter Paper II, we use the luminosities for individual sources found in this paper to discuss their disk stability and evolution.

\section{Data Analysis}

\subsection{Sample selection}

As of January 2013, 13 galactic UCXBs have reliable orbital period measurements (some tentative, see Table 1). Four other likely UCXBs show strong indications of their ultracompact nature, principally their X-ray/optical flux ratio \citep{Bassa06}, lack of H lines in their spectra \citep{Nelemans04,Nelemans06}, and burst characteristics \citep{Galloway10}.  Eight objects have been suggested to be UCXBs based principally on their low ($<$2 \% of Eddington) persistent X-ray luminosity \citep{intZand07}.   To study the UCXB X-ray luminosity function, we choose to omit the 8 objects suggested on this basis alone, as this method of identification would clearly bias the derived luminosity function.  We also note that a counter-example to the in't Zand method--an apparently persistent system below 2 \% of its Eddington luminosity, not edge-on, and with a 2.15 hour orbital period \citep{Engel11}--is now known, H1825-331 in the globular cluster NGC 6652.   Whether this system is a transient or not is debatable \citep{Verbunt95,Deutsch98}, but if it is a transient, it has been active as long as many systems thought to be persistent, so the discrepancy with the in't Zand method remains.

We cannot be certain that other identification methods (which are not uniform) do not introduce luminosity biases, but at least they are not obviously biased.
We list the remaining 17 systems in Table 1, separating them into transient and persistent systems.  Of these 17, 5 accreting millisecond pulsars are known transients, while the rest are (so far) deemed persistent. 

From the candidate ultracompacts listed by \citet{intZand07} and \citet{Nelemans10}, we exclude 4U 1822-00 \citep[due to its optical modulation period of 191 minutes,][]{Shahbaz07}, NGC 6652 A = H 1825-331 \citep[due to its 2.15 hr optical period,][]{Engel11}, NGC 6652 B (due to its apparently main-sequence donor \citealt{Heinke01}, and recent evidence against the proposed 45-minute period \citealt{Engel11}), and the $\omega$ Cen qLMXB \citep[due to its strong H$\alpha$ emission,][]{Haggard04}.  We also omit 4U 1905+000, due to a lack of high-cadence sensitive monitoring data (it has been quiescent since at least 1992; \citealt{Juett05}).

\subsection{Data selection \& analysis} 

We took our data from all-sky monitoring surveys wherever feasible.   
The Monitor of All Sky X-ray Image (\maxi) detector \citep{Matsuoka09, Sugizaki11} aboard the International Space Station (ISS) provides high-sensitivity lightcurves for bright X-ray binaries anywhere in the sky since August 2009, while the Rossi X-ray Timing Explorer's ({\it RXTE}) Proportional Counter Array (\pca, \citealt{Jahoda96}) scans of the Galactic Bulge \citep{Swank01} permit higher-sensitivity lightcurves of X-ray binaries in the Galactic Bulge from 1999 to 2012 (and expanded regions for shorter times).  

We obtained daily binned \maxi\ Gas Slit Camera data, in units of counts s$^{-1}$ cm$^{-2}$ in the 2.0-4.0 keV and 4.0-10.0 keV energy ranges, from the \maxi\ website\footnote{http://maxi.riken.jp/top/}.  We chose to omit the 10-20 keV energy range, due principally to its poorer signal-to-noise ratio, but also because we are focused on comparing to Chandra detections of extragalactic X-ray binaries below 10 keV. 
 \pca\ bulge scan data were obtained from C. Markwardt's webpage\footnote{http://lheawww.gsfc.nasa.gov/users/craigm/galscan/main.html}
 in the 2-60 keV band.
\maxi\ data collection began on MJD 55058 (2009/8/15), with our data spanning approximately 2 years and 8 months, while there is over 12 years worth of \pca\  bulge scan data for most sources (the bulge scans were extended to cover larger regions later). 

We used \maxi\ data for persistent sources outside the bulge scan region,  combining the 2-4 and 4-10 keV data and errors.  We discarded roughly 30 data points per source corresponding to dates the shuttle was docked at the ISS, and high data points that were clearly attributable to Sun glints (identifiable as a bright, elongated source moving over the \maxi\ map around the source over a few days).  We also identified data with large errors as noisy; the value of the error cutoff was determined on an individual basis for each source.  This amounted to  $\sim$0.5--10\% of the data from each source.  Combined \maxi\ datapoints less than 3 $\sigma$ above zero were considered upper limits; we create 3 $\sigma$ upper limit points by replacing the datapoint with an upper limit three times the error above zero.
For \pca\ data, we used a 4 $\sigma$ detection limit, since the \pca\ is more sensitive, and the crowded bulge leads to increased systematic errors.  For systems known to be transients (e.g. observations with sensitive X-ray detectors found them in quiescence), upper limits were taken to indicate quiescence, and were simply replaced with a value of 0. Persistent sources were essentially always detected in \pca\ data.

Many of these UCXBs exhibit substantial variability. The lightcurves of 4U 1850-087, 4U 1728-34, and 2S 0918-549 (Fig. 1) demonstrate variation of up to an order of magnitude, on timescales from days to months.  Such variability was highlighted recently by \citet{Maccarone10} for 4U 1543-40 (the UCXB in NGC 1851).  Similar variability was noted from, e.g. the UCXBs 4U 1915-05 by \citet{Simon05} and 1A 1246-588 by \citet{intZand08}; see below. 

We converted the \maxi\ data into 2.0 -- 10.0 keV intrinsic luminosities using the Crab nebula as a standard calibration source.  In the 2-10 keV energy band, an average photon flux of 3.12 cts s$^{-1}$ cm$^{-2}$ was found for the Crab Nebula.
Using the Crab's flux in the 2.0-10.0 keV range ($2.16\times10^{-8}$ ergs cm$^{-2}$ s$^{-1}$), we find $6.9\times10^{-9}$ ergs (2-10 keV)/ct.  We assume that our sources have roughly Crablike spectra (power-law of photon index 2.1) within this band.  Altering the photon index by 1 in either direction, or increasing $N_H$ up to ten times larger, alters the unabsorbed flux by at most 30\% \citep{intZand07}.  
 For the \pca\ data, the Portable Interactive Multi-mission Simulator (PIMMS)\footnote{http://asc.harvard.edu/toolkit/pimms.jsp} was used to convert photon fluxes (counts per s per 5 PCU) into unabsorbed energy fluxes in the 2.0 - 10.0 keV range assuming a photon index of 2 and the Galactic $N_H$ values in Table 1. 
  
We compute 2-10 keV intrinsic luminosities using the best-estimate distances in Table 1.  Many of these distances come from photospheric radius expansion bursts, or from estimates of globular cluster distances, and are thus well-constrained.  Other distances--for 4U 1543-624, 4U 1626-67 and four of the five transient systems--are relatively poorly constrained (though the location of several of the transients near the Galactic center suggests a location in the Galactic bulge).  The only distance estimate for 4U 1543-624 comes from assuming that its $\dot{M}$ is powered by gravitational radiation, and that it follows standard white dwarf UCXB evolution.
  
\clearpage

\begin{figure}
\figurenum{1}
\begin{center}$
\begin{array}{cc}
\includegraphics[scale=0.45]{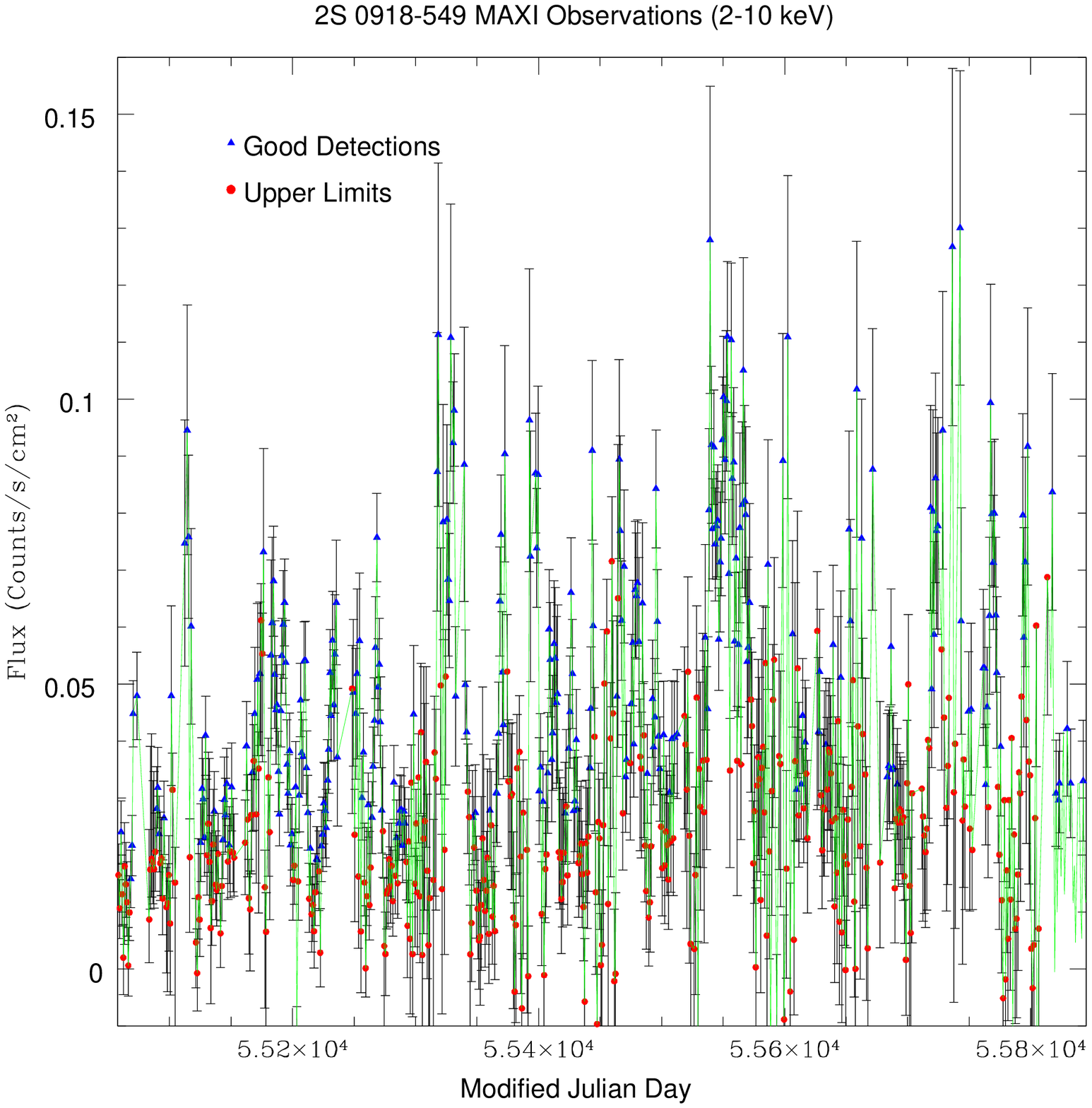} 
\includegraphics[scale=0.45]{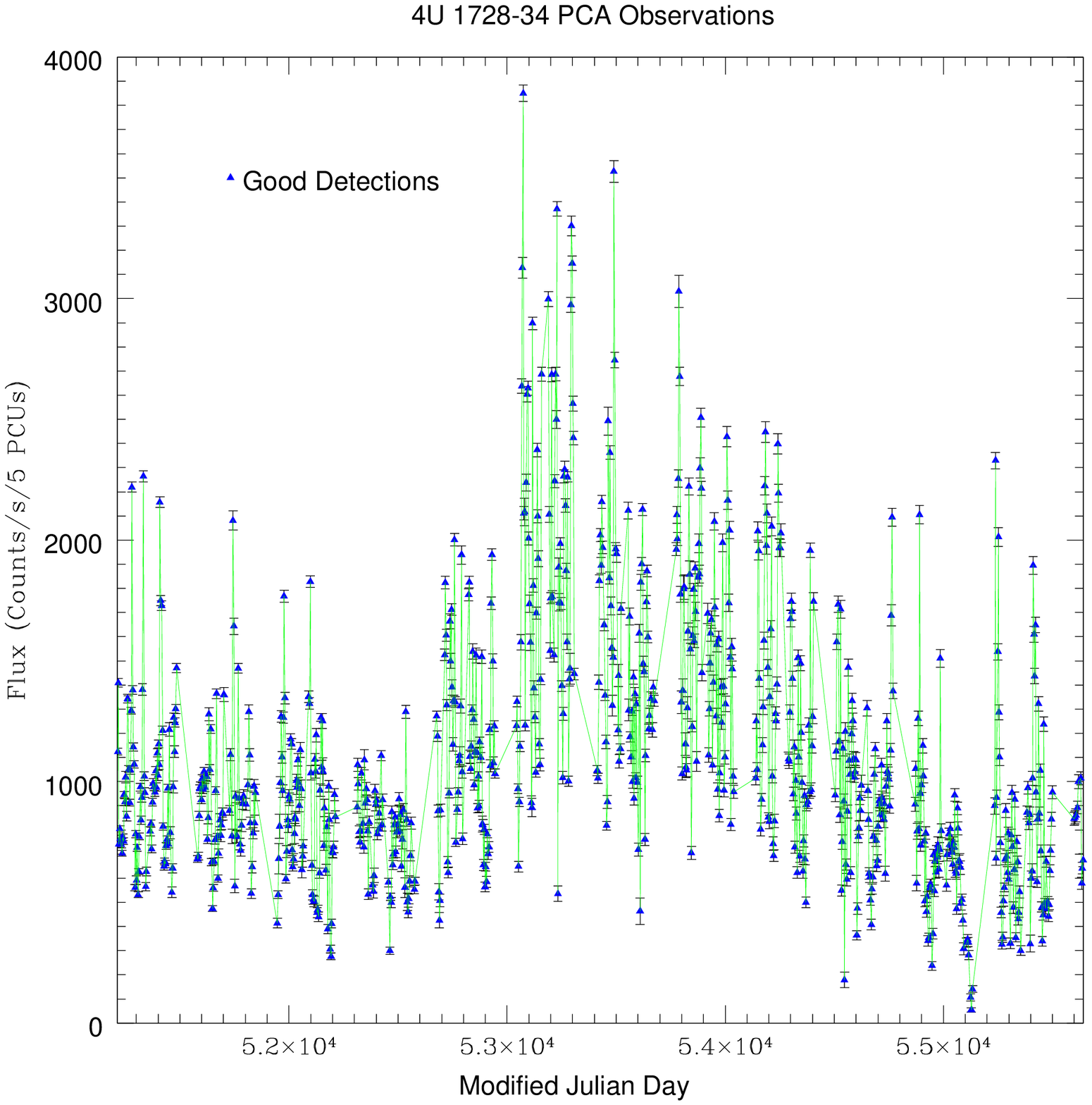}
\end{array}$
\includegraphics[scale=0.45]{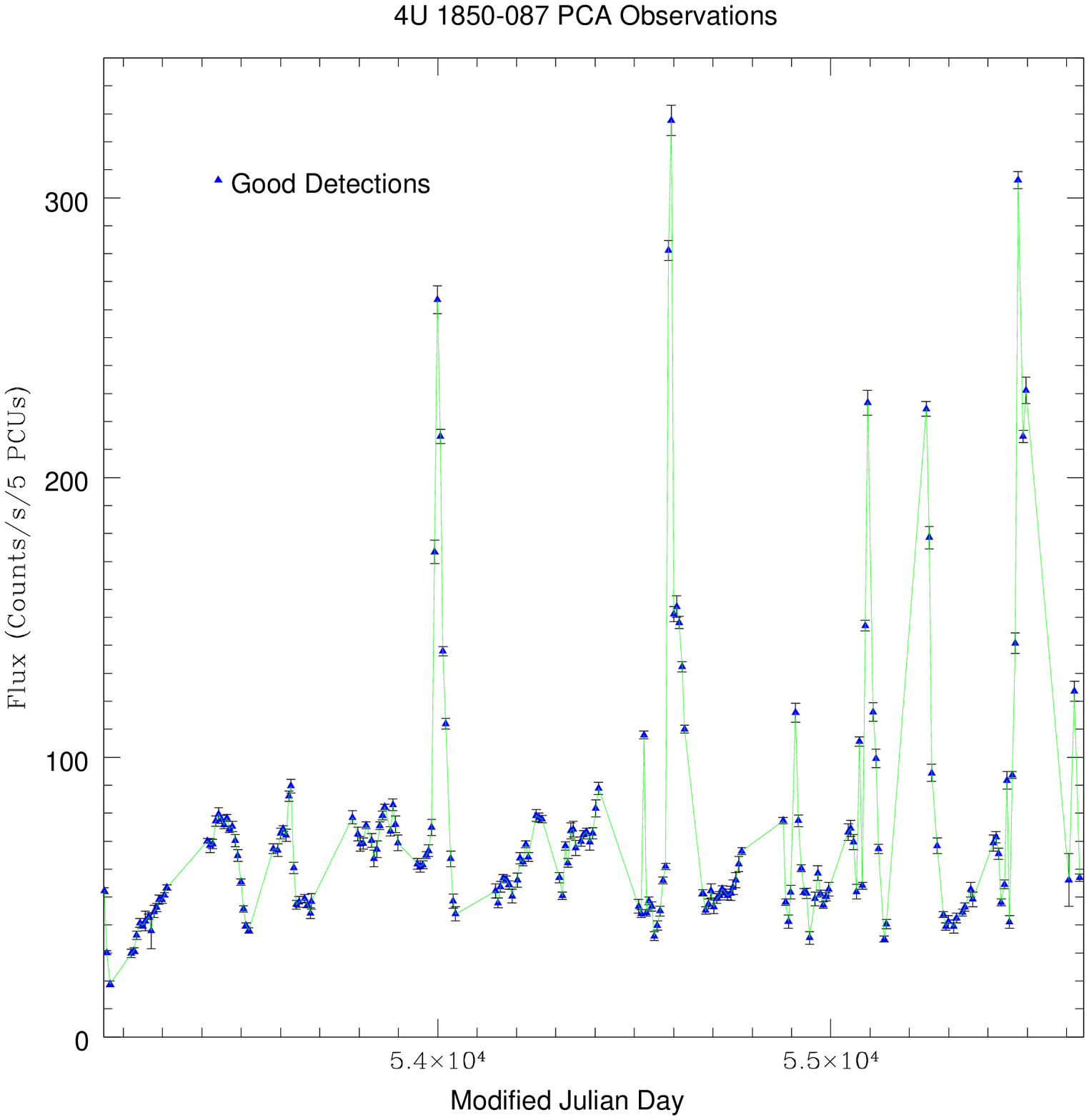}
\caption{ \maxi\ and \pca\ lightcurves of some persistent UCXB systems showing strong flaring; 2S 0918-549 (\maxi), 4U 1728-34 (\pca), and 4U 1850-087 (\pca).}
\end{center}
\label{lightcurve}
\end{figure}
\clearpage
\begin{deluxetable}{ ccccccccc }
\tablecaption{\textbf{UCXB Information}}
\tablehead{  Source & Location & Distance & Period & $N_H$  & \multicolumn{2}{c}{MAXI points} & \multicolumn{2}{c}{PCA points}   \\
    & & (kpc) & (mins) & ($10^{21}$ cm$^{-2}$) & $N_{Good}$ &  $N_{Limits}$ & $N_{Good}$ & $N_{Limits}$ }
\startdata
\multicolumn{9}{c}{Persistent systems} \\
  \hline
 4U 1728-34 & Field & $5.2\pm0.8^{a}$ & $10.8?^{a}$  & $22.9^{a}$ & - & - & 803 & 0 \\
  4U 1820-303 & GC& $7.9\pm0.4^{b}$ & $11^c$ &  1.6$^d$  & 600 & 0    &  2013 & 0 \\
  4U 0513-40 & GC & $12.1\pm0.6^b$ & $17^e$  &  0.26$^d$  & 144 & 507    &  - & - \\
  2S 0918-549 & Field & $5.4\pm0.8^f$ & $17.4^g$ & 3.0$^h$ & 300 & 519  & - & - \\
  4U 1543-624 & Field & $7.0^i$ & $18.2^i$ & 3.5$^h$ & 667 & 50 &  - &  - \\
 4U 1850-087 & GC & $6.9\pm0.3^b$ & $20.6^j$ & 3.9$^d$ & 106 & 645  & 195 & 0  \\
  M15 X-2 & GC & $10.4\pm0.5^b$ & $22.6^k$ & 0.67$^l$ & - & - & - & -  \\
 4U 1626-67 & Field & $8^{+5\ }_{-3}$ $^m$ & $42^m$ & 1.4$^n$ & 739 & 108 & -  & - \\
 4U 1916-053 & Field & $9.3\pm1.4^o$ & $50^p$ &  3.2$^q$ & 332 & 453  & -  & - \\
 4U 0614+091 & Field & $3.2\pm0.5^r$ & $51?^s$ & 3.0$^t$ & 567 & 50  &  - &  -  \\
 1A 1246-588 & Field & $4.3\pm0.6^u$ & ?  & 2.5$^u$ & 122 & 389  & - & -\\
 4U 1812-12 & Field & $4.6\pm0.7^v$ & ?      & $15.0^w$ & - & - & 405 & 0 \\
 \hline %
  \multicolumn{9}{c}{Transient systems} \\
  \hline
  XTE\_J1807-294 & Bulge & $8^{+4}_{-3.3}$ $^x$ & $40.1^y$ & $5.6^{z}$ & 105 & 602 & 35 & 775 \\
 XTE\_J1751-305 & Bulge & $8^{+0.5\ }_{-1.3}$ $^{\alpha}$ & $42^{\beta}$ & $9.8^{\gamma}$ & - & - & - & - \\
  XTE J0929-314 &  Field & $8^{+7\ }_{-3}$ $^x$ &  43.6$^{\delta}$ & 0.76$^{\epsilon}$ & - & - & - & \\
 Swift J1756.9-2508 & Bulge & $8\pm4^{\zeta}$ & 54.7$^{\zeta}$ &  54$^{\zeta}$ & - & - & 10 & 727 \\
 NGC 6440 X-2 & GC & 8.5$\pm0.4^b$ & 57.3$^{\eta}$ & 5.9$^b$ & & & 36 & 785  \\
\enddata
\tablecomments{Known and suspected UCXBs in our sample, with best estimates of their distance, period and $N_H$, plus the number of usable data points from {\it MAXI} and the {\it PCA} bulge scans. Location in the Galactic field, (direction of the) bulge, or in a Globular Cluster (GC) is also specified.
Errors are ranges from indirect estimates; 15\% errors on bursts \citep{Kuulkers03}; 5\% errors on GC distances.  Periods supported by only weak evidence have '?'s. 
 References: $^{a}$\citet{Galloway10}; $^{b}$\citet{Harris10}; $^{c}$\citet{Stella87}; $^{d}$\citet{Sidoli01}; $^{e}$\citet{Zurek09}; $^{f}$\citet{intZand05b}; $^{g}$\citet{Zhong11}; $^{h}$\citet{Juett03b}; $^{i}$\citet{Wang04}, distance estimate assumes $\dot{M}$ driven by gravitational radiation; $^{j}$\citet{Homer96}; $^k$\citet{Dieball05}; $^l$\citet{White01}; $^m$\citet{Chakrabarty98}; $^n$\citet{Krauss07}; $^o$\citet{Yoshida93}; $^p$\citet{Walter82}; $^q$\citet{Church98}; $^r$\citet{Brandt92}; $^s$\citet{Shahbaz08}; $^t$\citet{Piraino99}; $^u$\citet{intZand08}; $^v$\citet{Bassa06}; $^w$\citet{Tarana06}; $^x$\citet{Galloway06b}; $^y$\citet{Markwardt03}; $^z$\citet{Falanga05}; $^{\alpha}$\citet{Papitto08}; $^{\beta}$\citet{Markwardt02}; $^{\gamma}$\citet{Miller03}; $^{\delta}$\citet{Galloway02}; $^{\epsilon}$\citet{Juett03}; $^{\zeta}$\citet{Krimm07}; $^{\eta}$\citet{Altamirano10}. 
}
\label{all_sources}
 \end{deluxetable}
 \clearpage
For four faint transients (XTE J1751-305, NGC 6440 X-2, XTE J0929-314, and Swift J1756.9-2508) and M15 X-2, neither \maxi\ nor \pca\ bulge scan lightcurves provide a reliable history.  For XTE J1751-305, we create daily 2-10 keV flux lightcurves of the 2002, 2005, 2007 and 2009 outbursts from the literature \citep{Markwardt02, Grebenev05, Swank05, Falanga07, Markwardt07, Riggio11}, and take other PCA bulge scan measurements as evidence of quiescence.  
For NGC 6440 X-2, we use the \pca\ bulge scan observations as for other sources (as it was detected in several, \citealt{Heinke10,Patruno10b}), but exclude the three major outbursts, which were all produced by the other known transient in that cluster, SAX J1748.9-2021 \citep{Patruno10,Altamirano08}.  As it is a transient, we only use detections, and set upper limit observations to indicate quiescence.  For XTE J0929-314, we use the outburst lightcurve data from \citet{Galloway02}.  It was clearly observed by the \RXTE\  All-Sky Monitor (ASM)\footnote{http://heasarc.gsfc.nasa.gov/docs/xte/asm\_products.html} in its 2002 outburst, but was not otherwise detected by \RXTE\ ASM, so we consider the remaining ASM history since 1996 to be a record of quiescence.  For Swift J1756.9-2508, we use the 2007 outburst lightcurve data from \citet{Krimm07} along with the \pca\ bulge scans, converting the {\it Swift} Burst Alert Telescope (BAT) 15-50 keV fluxes to 2-10 keV fluxes using a typical flux ratio of 3 (which aligns the BAT and \pca\ flux estimates when they are simultaneous). 
The 2007 outburst was missed by the \RXTE\ \pca\ bulge scans due to Sun constraints, while other outbursts during the bulge scan era have been detected, so we take bulge scan upper limits as evidence of quiescence.

 M15 X-2 cannot be separated from AC 211 (X 2127+12) with any X-ray instrument but Chandra.  We use one archival  and six reported \Chandra\ observations of M15 to interpret its long-term lightcurves. \citet{White01}'s \Chandra\ HETGS spectral fit gives $L_X$(2-10 keV)$=9\times10^{35}$ ergs/s.  Applying this spectral fit to three \Chandra-HRC observations in 2001 \citep{Hannikainen05} and one in 2007 \citep{Heinke09b} gives $L_X=1.0\pm0.1\times10^{36}$ ergs/s in each case. We extract M15 X-2's readout streak spectrum (using the standard analysis recipe \footnote{http://cxc.harvard.edu/ciao/threads/streakextract/}) from archival \Chandra\ ACIS-S ObsID 11029, finding $L_X=7\times10^{35}$ ergs/s. Finally, \citet{Sivakoff11} report that M15 X-2 is responsible for a major X-ray brightening of M15 in 2011, giving $L_X=1.1\times10^{37}$ ergs/s. We thus suggest that M15 X-2 is responsible for similar previous large flares in M15 seen by the \RXTE\ ASM. M15 X-2 appears to be usually in the range of  $7\times10^{35}<L_X<1\times10^{36}$ ergs/s, except for the bright flares seen by \RXTE\ ASM and \maxi. AC 211 is typically fainter than M15 X-2, but more variable \citep{Hannikainen05}. 
Thus we take M15 X-2 to be between $7\times10^{35}<L_X<1\times10^{36}$ ergs/s when the total flux from M15 is below $2\times10^{36}$ ergs/s.  
We use \RXTE\ ASM data to get better long-term statistics of the bright flares (as only one, very bright flare was detected with MAXI). We take a 5-day average of \RXTE\ ASM datapoints, and select data with countrate $>1.5$ cts/s, and errors $<0.5$ cts/s, as ``flaring'', finding 68 such datapoints and 915 ``normal'' datapoints. Although the resulting luminosity function is overly simplistic, it seems to represent the key elements of M15 X-2's behavior reasonably well.

\subsection{X-ray luminosity functions}
Calibrated data points were arranged into thirty-one luminosity bins between 0 and 10$^{38}$ ergs/s, where all luminosity values less than 10$^{35}$ were placed in the lowest bin. The errors for each bin were computed using Gehrels' upper limit approximation, $1+\sqrt{N+0.75}$ \citep{Gehrels86}, and taken to be symmetric. X-ray luminosity functions (XLFs) for each source can be found in Fig. 2, with upper limits also plotted for \maxi\ sources. For the five transient sources considered, the XLFs in Fig. 2 represent periods of outburst only. 

We create empirical XLFs for the UCXB population in two ways: taking one observation for each source  creates a ``snapshot'' luminosity function of the UCXB population (as from a single observation of our Galaxy) and compiling 100 observations of each source creates a combined luminosity function over time.
  Both require random sampling from their intrinsic luminosity functions, incorporating our detections and upper limits.  For \pca-detected systems, we use the detections alone (Table 2). For known transients (Table 3), upper limits are assumed to indicate the source is in deep quiescence ($L_X<<10^{35}$ ergs/s), as is typically found by deep \Chandra\ or {\it XMM} observations. This assumption appears valid given both the rapid evolution of transient outbursts and recent results monitoring Galactic globular clusters for low-luminosity transients (Altamirano et al.\ 2013, in prep).

 For persistent sources with upper limit observations, we statistically estimate the true fluxes represented by upper limits using the maximum likelihood method of \citet{Avni80}.  This method uses an analytic, recursive function to estimate the `true' shape represented by input data and upper limits, and was designed to be used with relatively sparse and binned astronomical data.  Avni presents a formula appropriate for data with \textit{lower} limits; since in our case, we deal with \textit{upper} limits, we delineate a brief derivation of our treatment, modified from Avni, in Appendix A (see also \citealt{Feigelson85}).  We compute maximum likelihood X-ray luminosity functions for the persistent sources for which we have upper limits (Fig. 3, Table 4).

\clearpage

\begin{figure}
 \figurenum{2}$
\begin{array}{cc}
\includegraphics[scale=0.42]{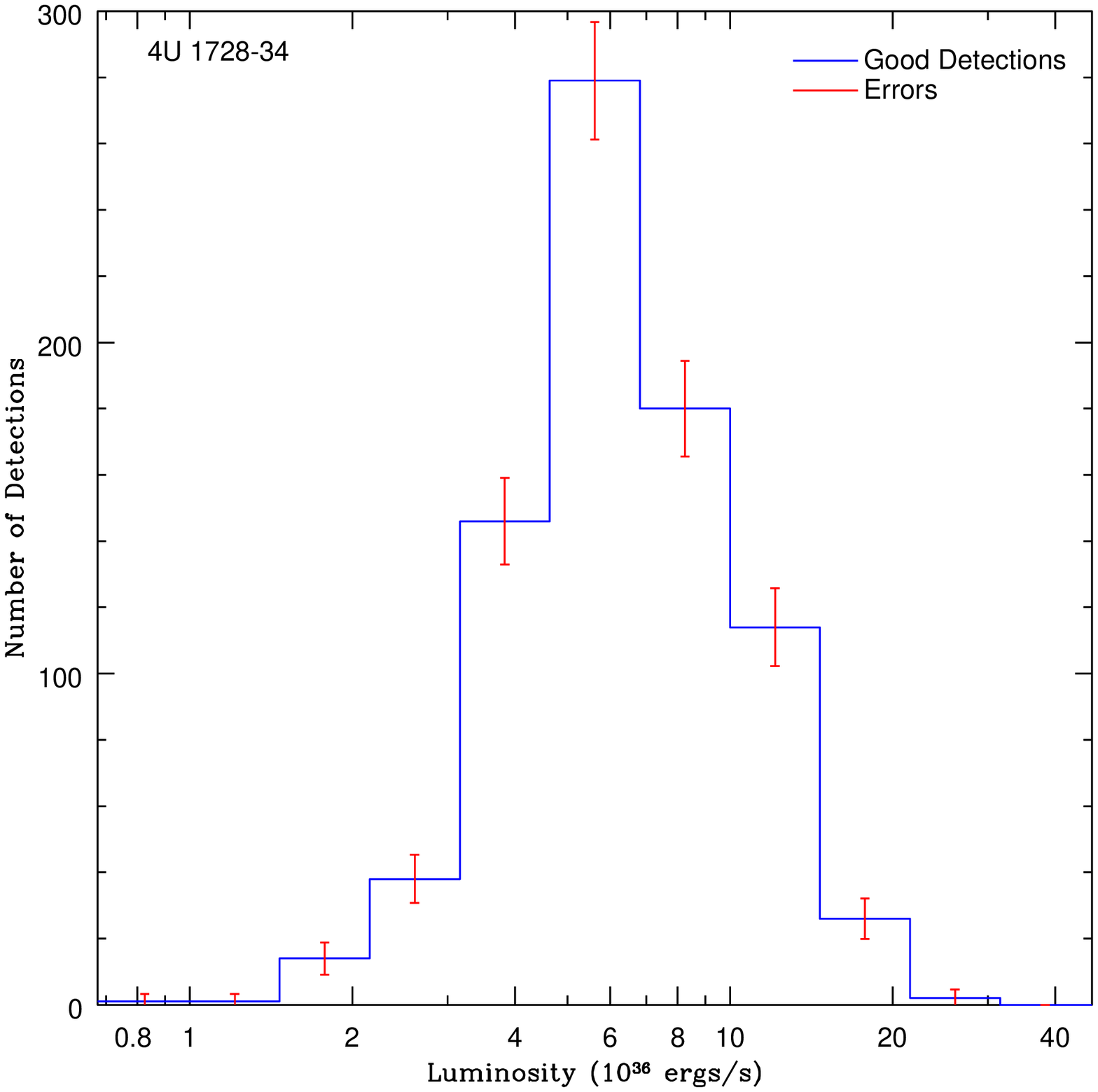} 
\includegraphics[scale=0.42]{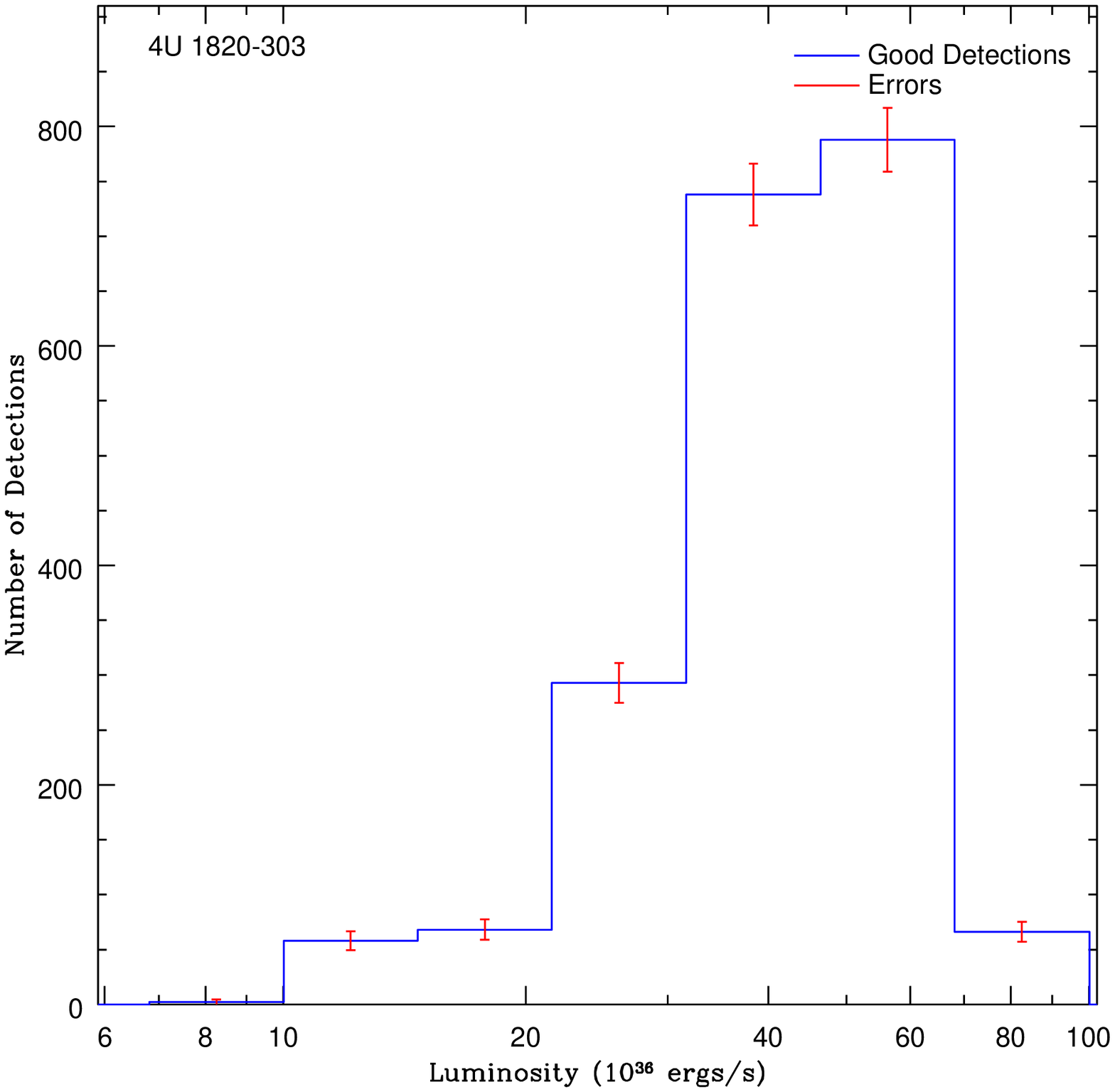} 
\end{array}{cc}$
\vspace{-1mm}$
\begin{array}{cc}
\includegraphics[scale=0.42]{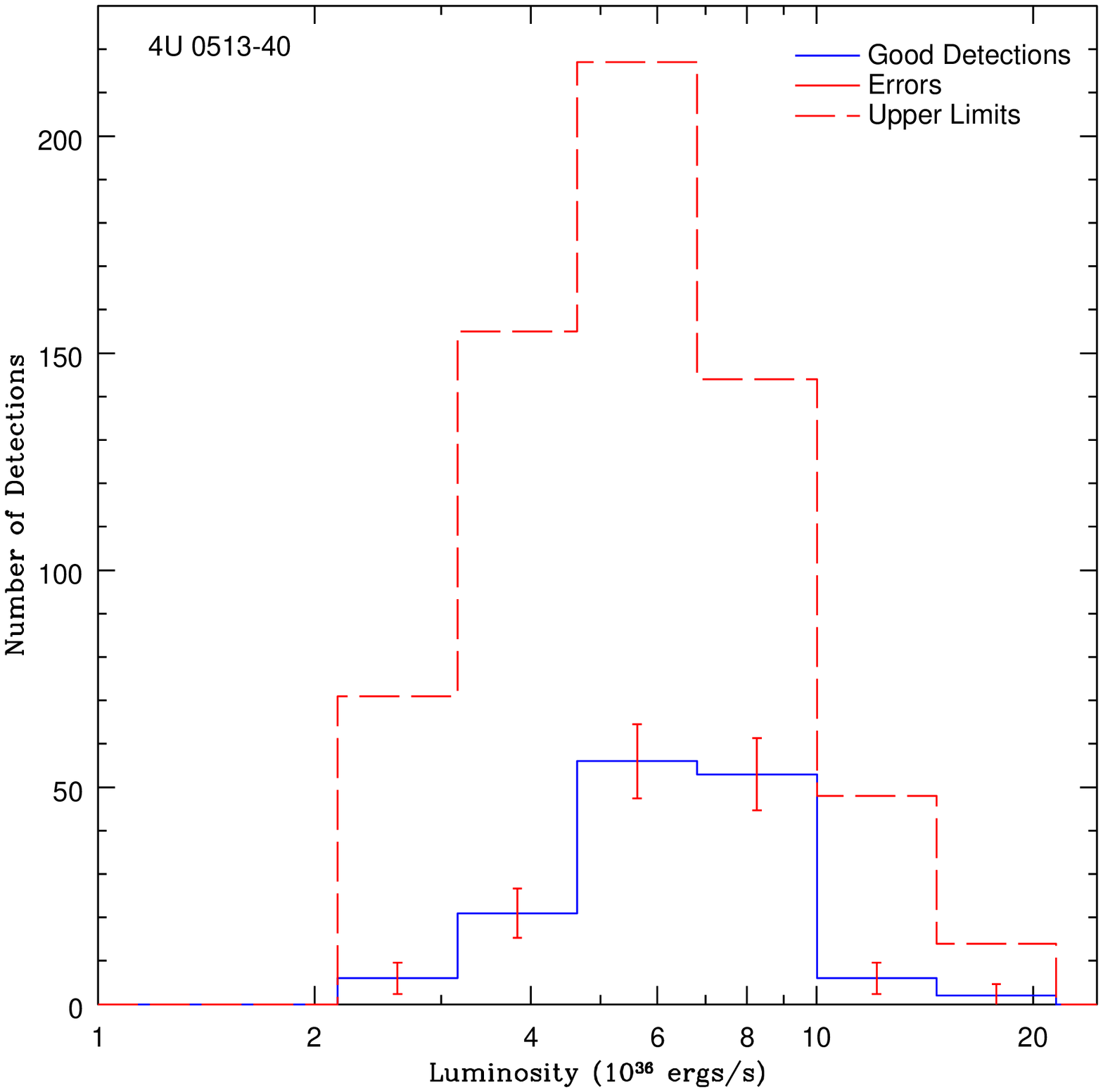} 
\includegraphics[scale=0.42]{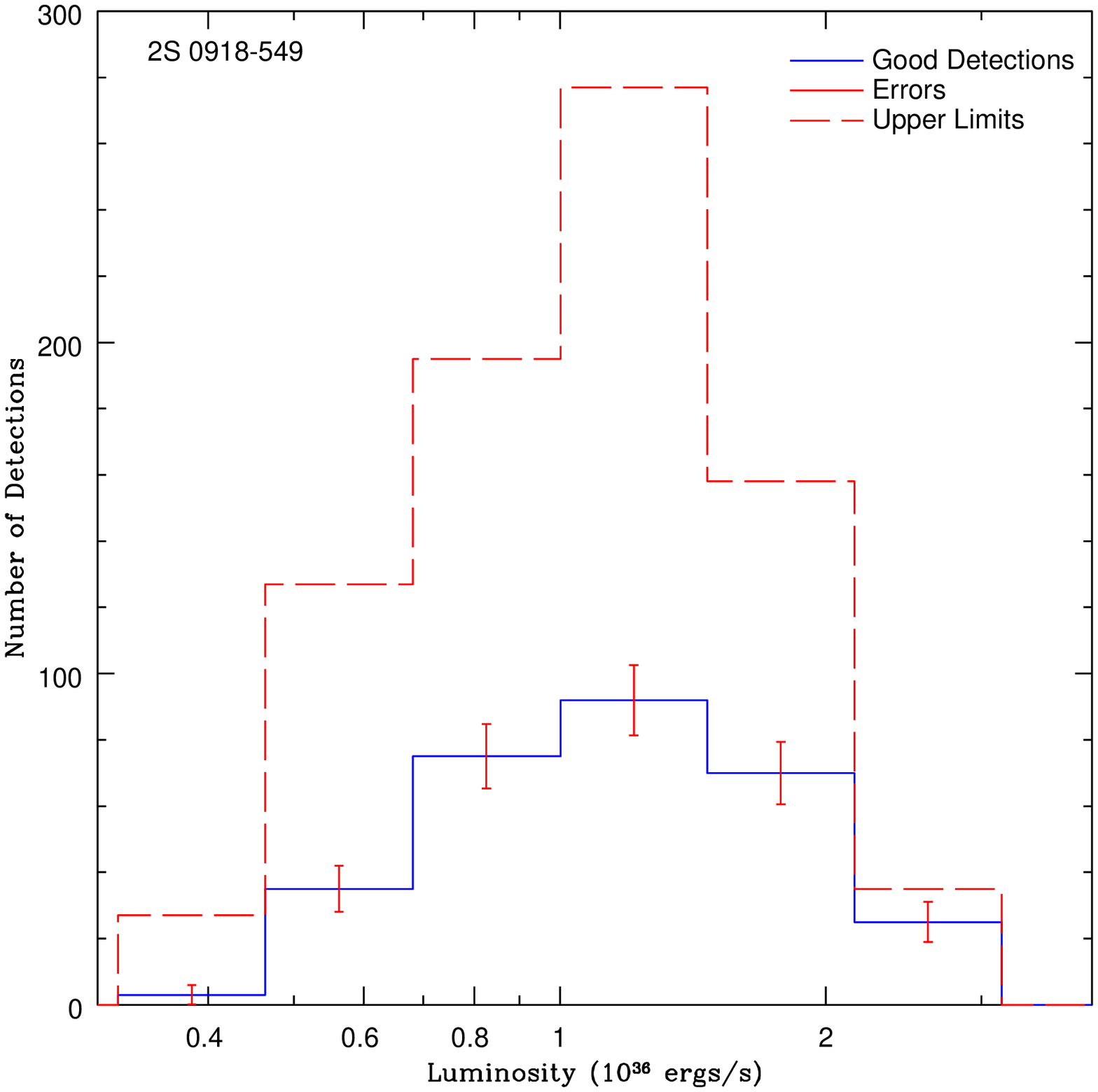} 
\end{array}{cc}$
   \caption{Observed XLFs for all 17 UCXB sources studied in this paper. XLFs of sources with both \maxi\ and \pca\ data are computed using the higher-quality \pca\ data. Upper limits are plotted (dashed red lines) where relevant (only for \maxi\ data).  For transient sources, upper limits are taken to indicate quiescence and a luminosity of 0, and are not plotted. }
\label{fig:cont}
\end{figure}
\clearpage

\begin{figure}
 \begin{center}
 \figurenum{2}$
\begin{array}{cc}
\includegraphics[scale=0.42]{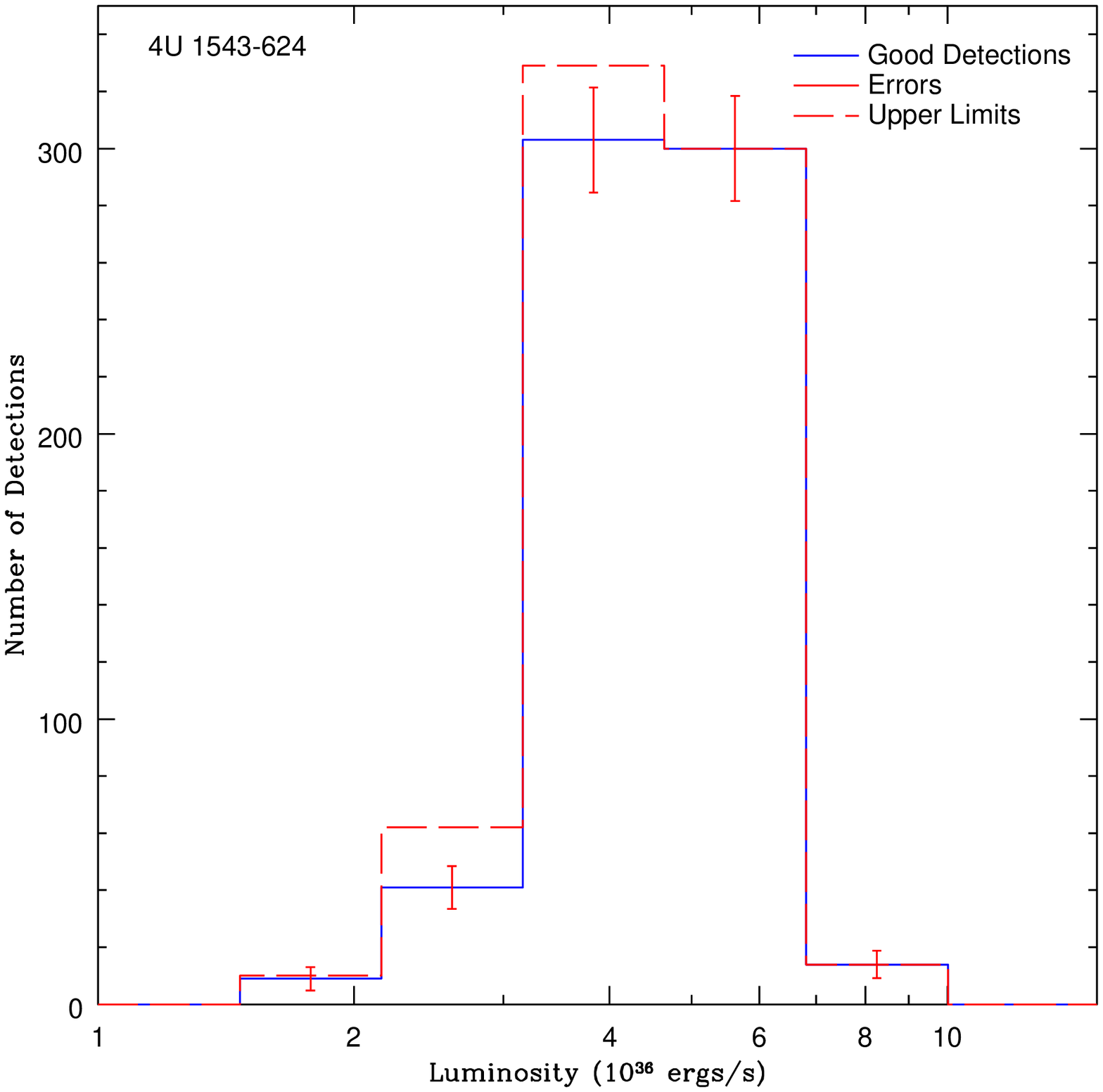}
\includegraphics[scale=0.42]{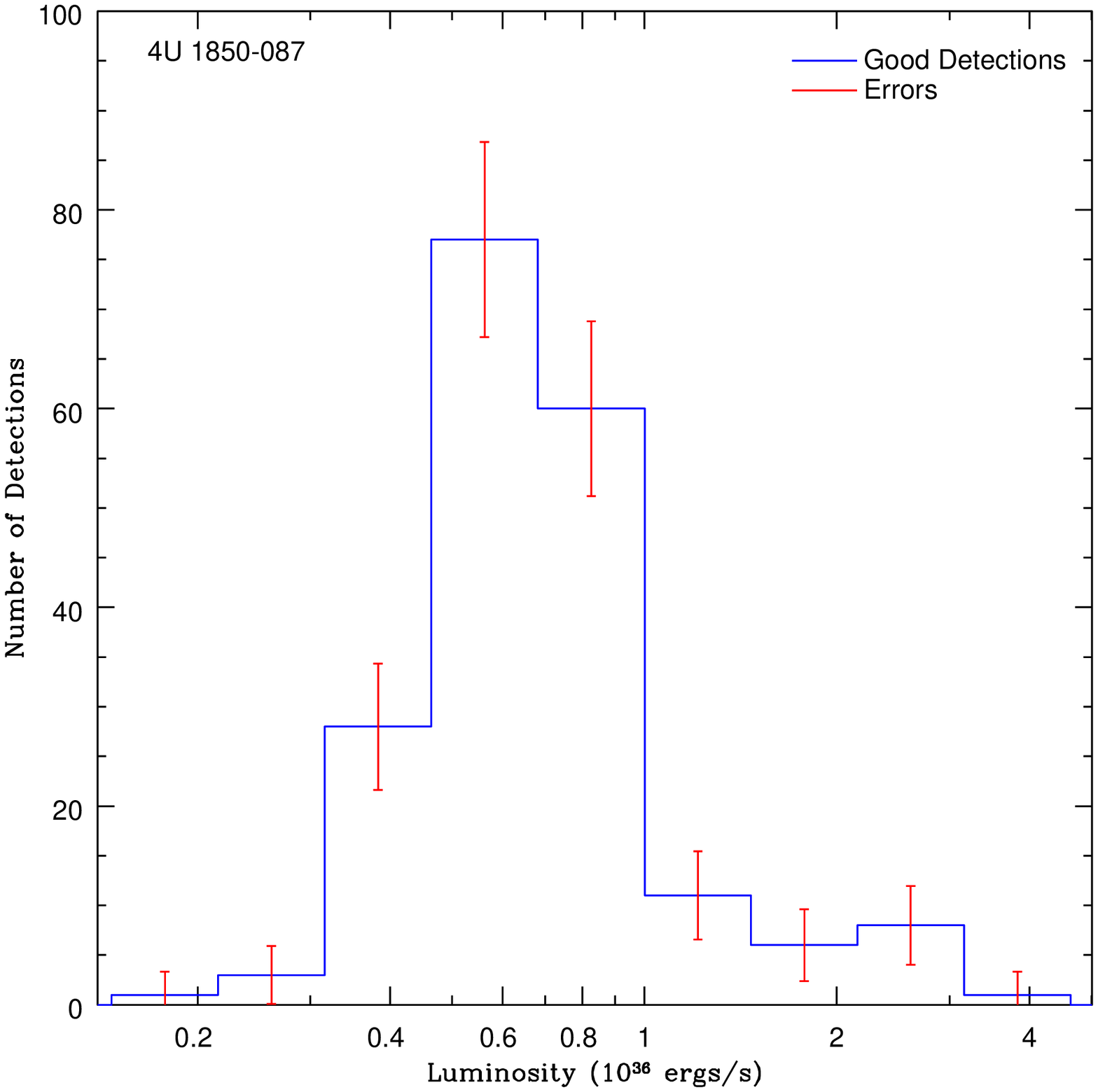} 
\end{array}{cc}$
\vspace{-4mm}$
\begin{array}{cc}
\includegraphics[scale=0.42]{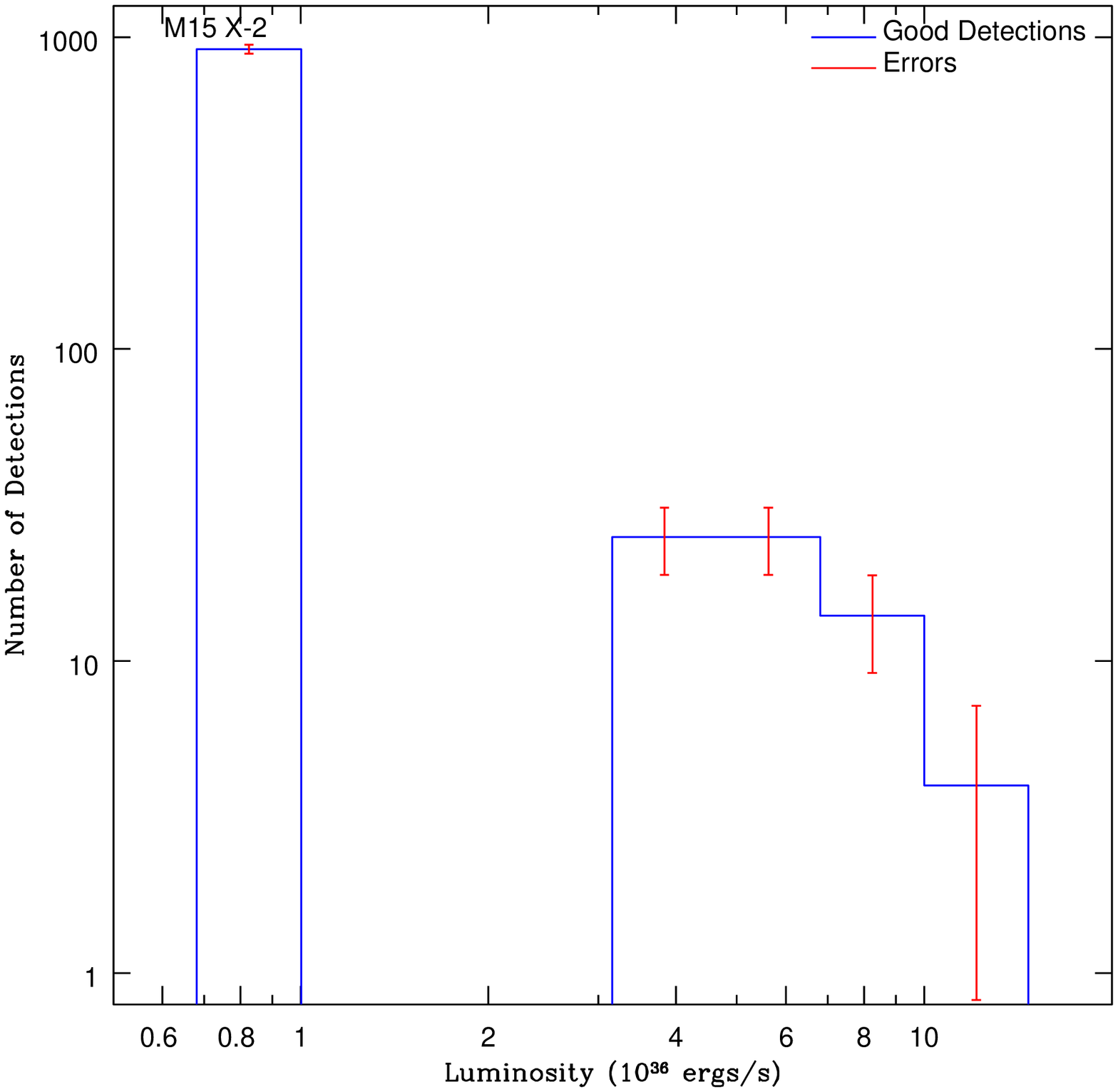} 
\includegraphics[scale=0.42]{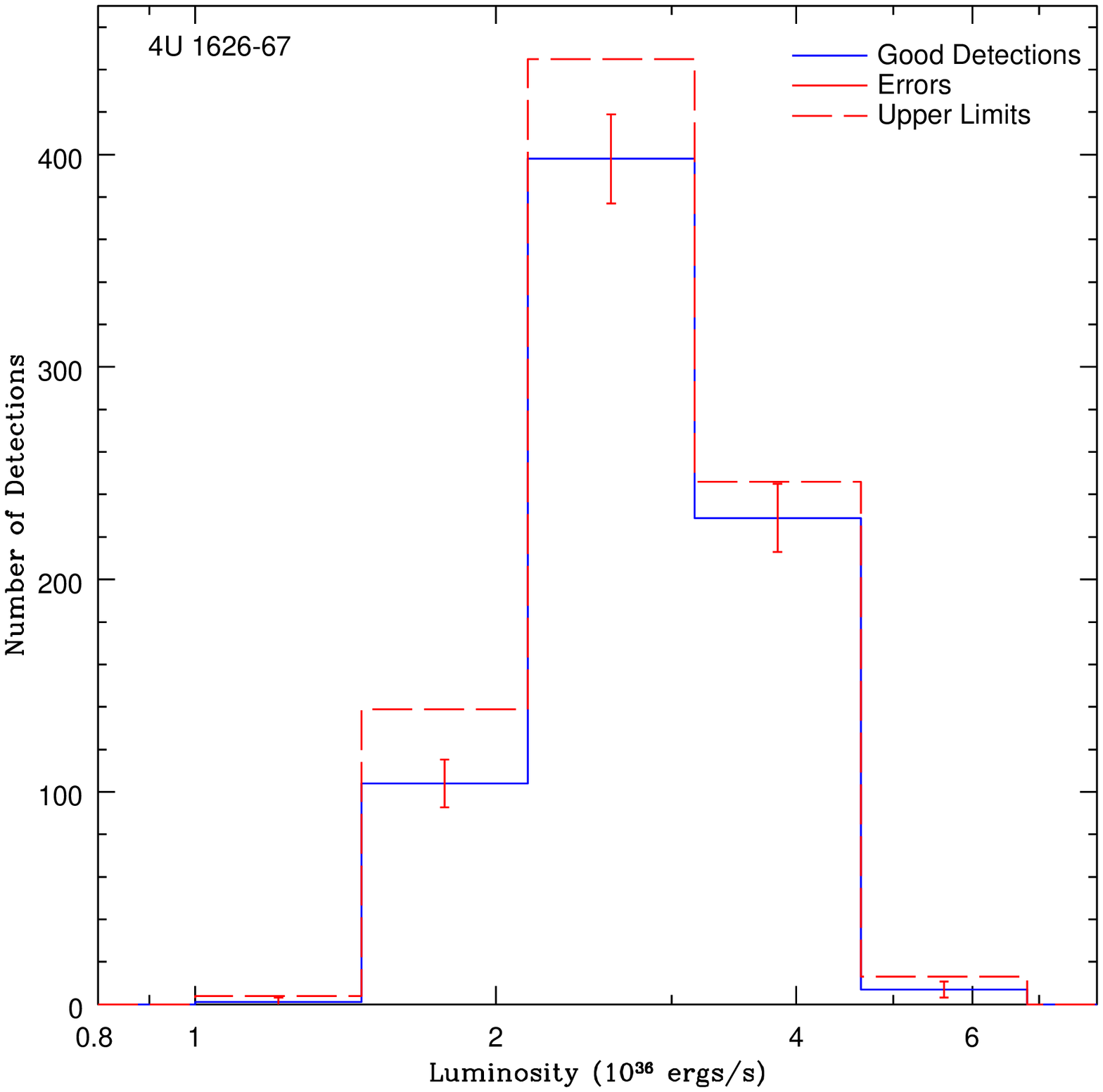} 
\end{array}{cc}$
\vspace{-4mm}$
 \begin{array}{cc}
\includegraphics[scale=0.42]{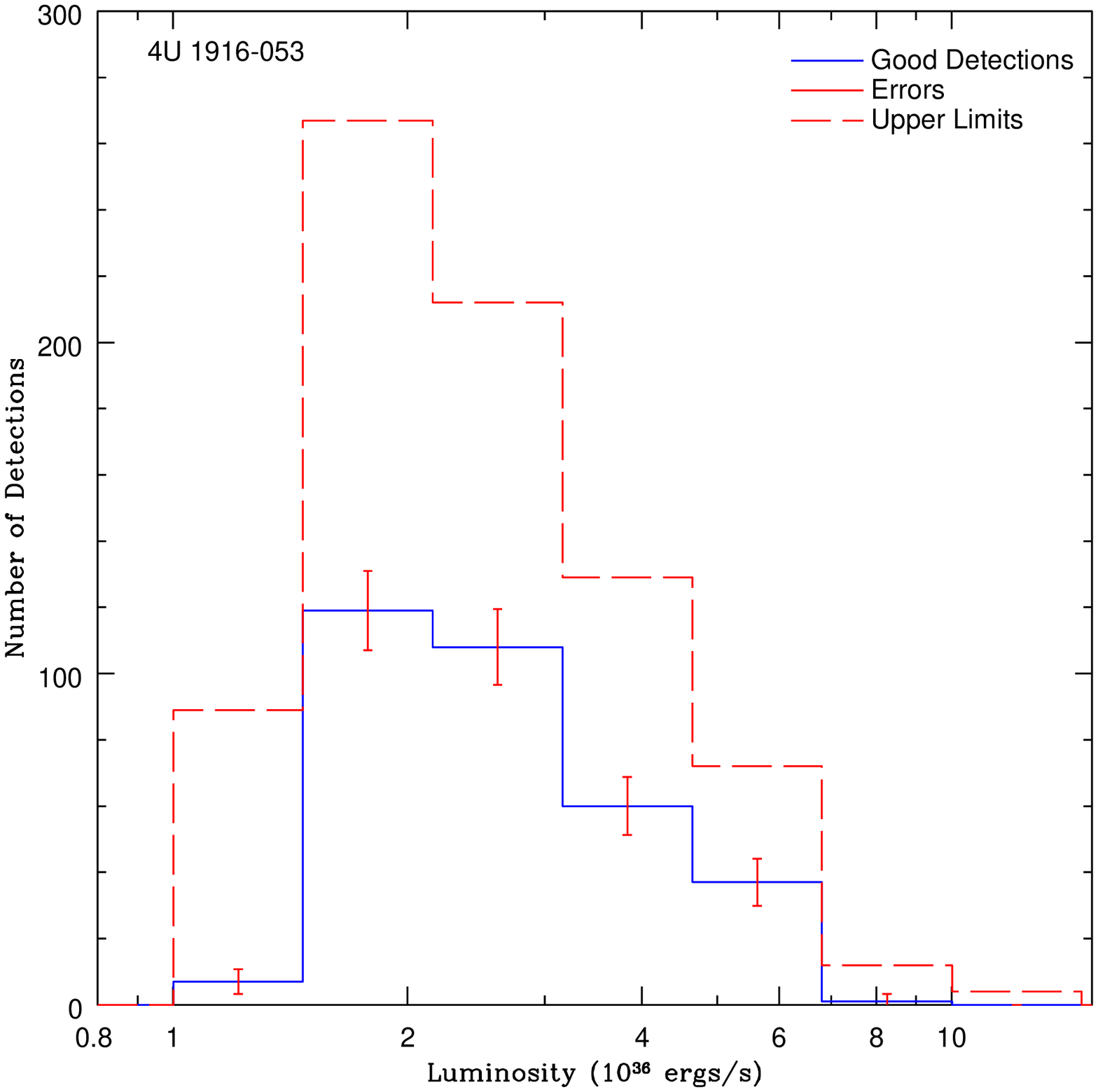} 
\includegraphics[scale=0.42]{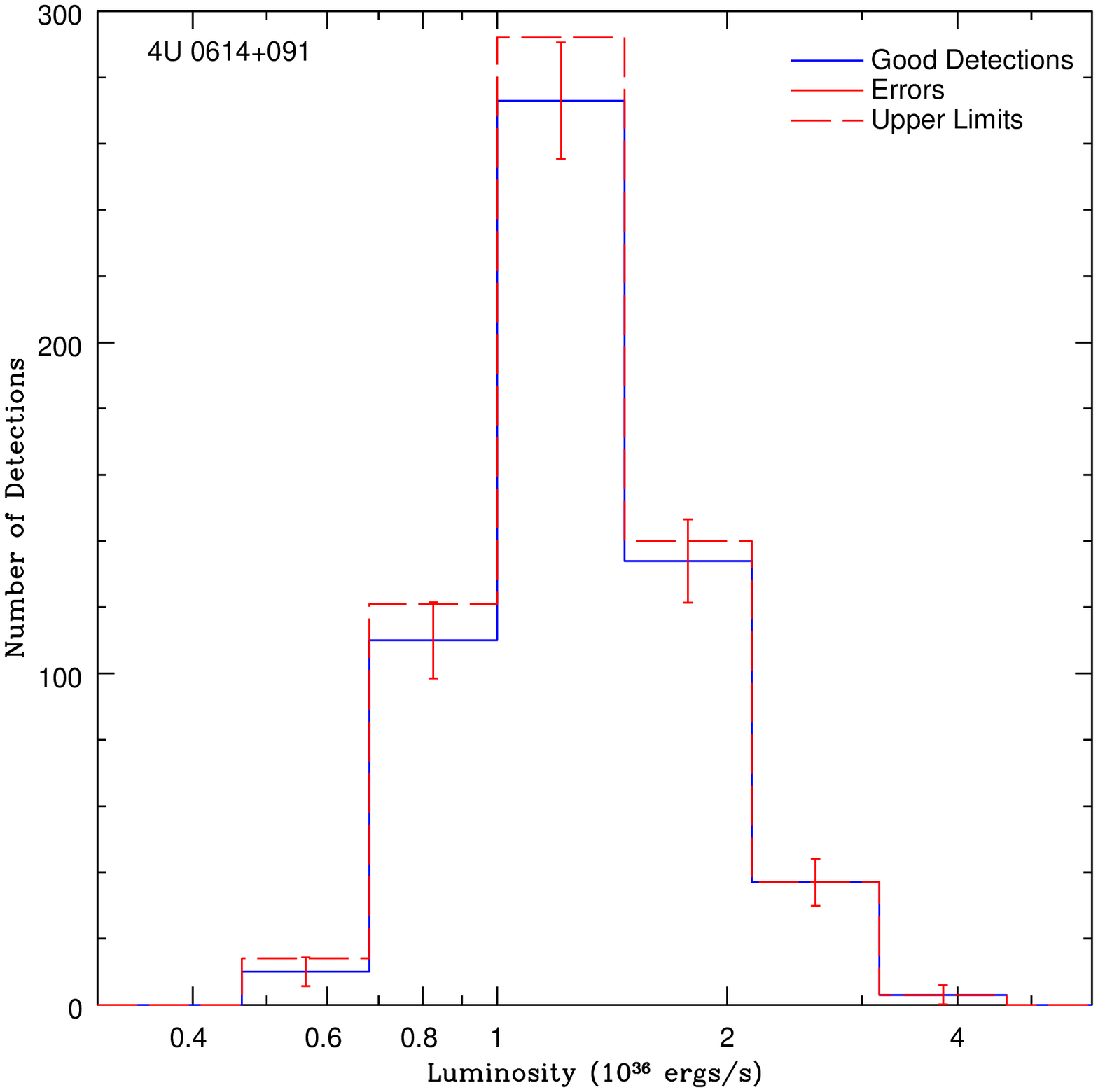} 
\end{array}{cc}$
\addtocounter{figure}{-1}
  \caption[]{cont'd. }
\label{fig:cont2}
\end{center}
\end{figure}
\clearpage

\begin{figure}
 \begin{center}
 \figurenum{2}$
\begin{array}{cc} 
\includegraphics[scale=0.42]{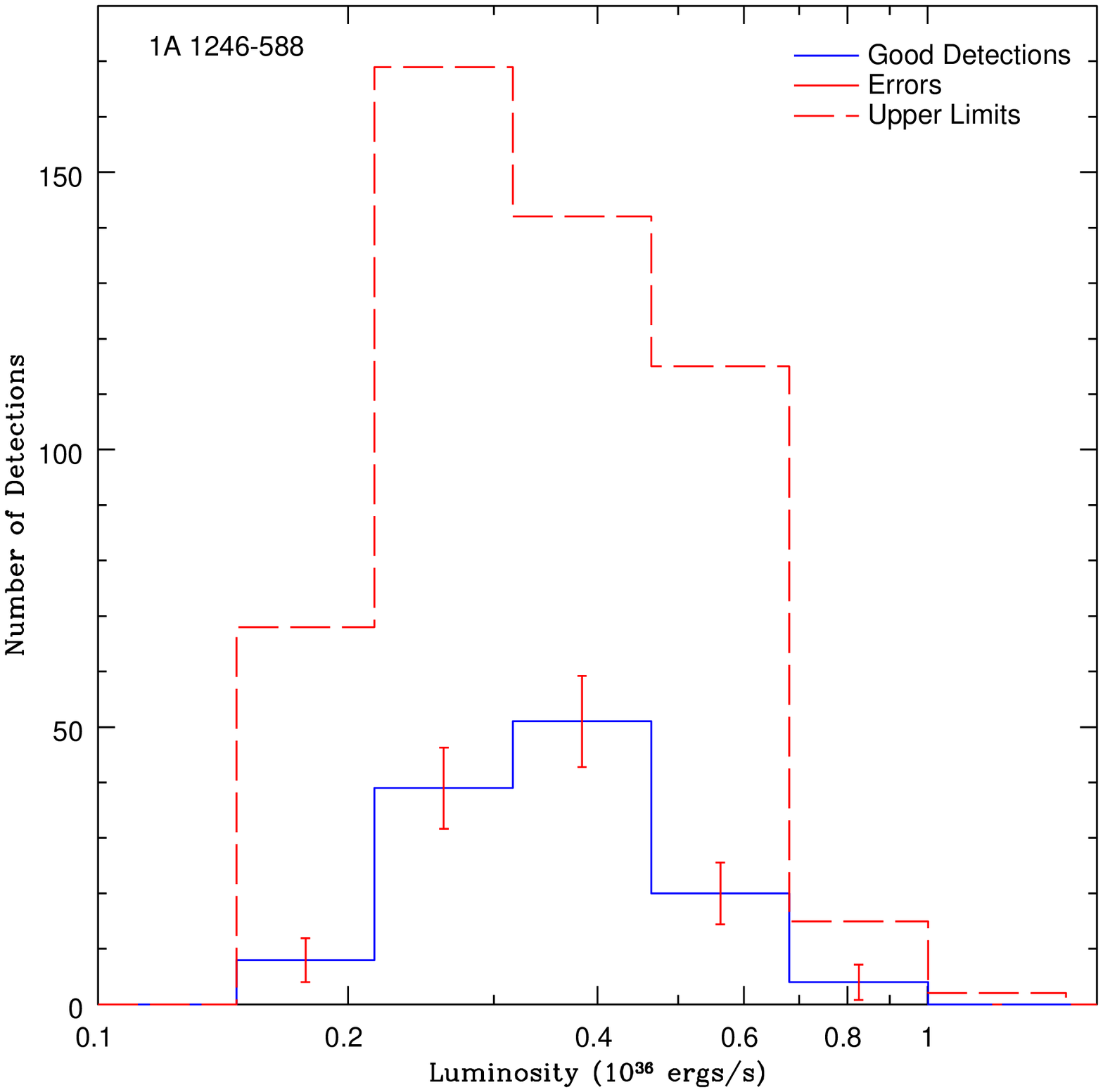}
\includegraphics[scale=0.42]{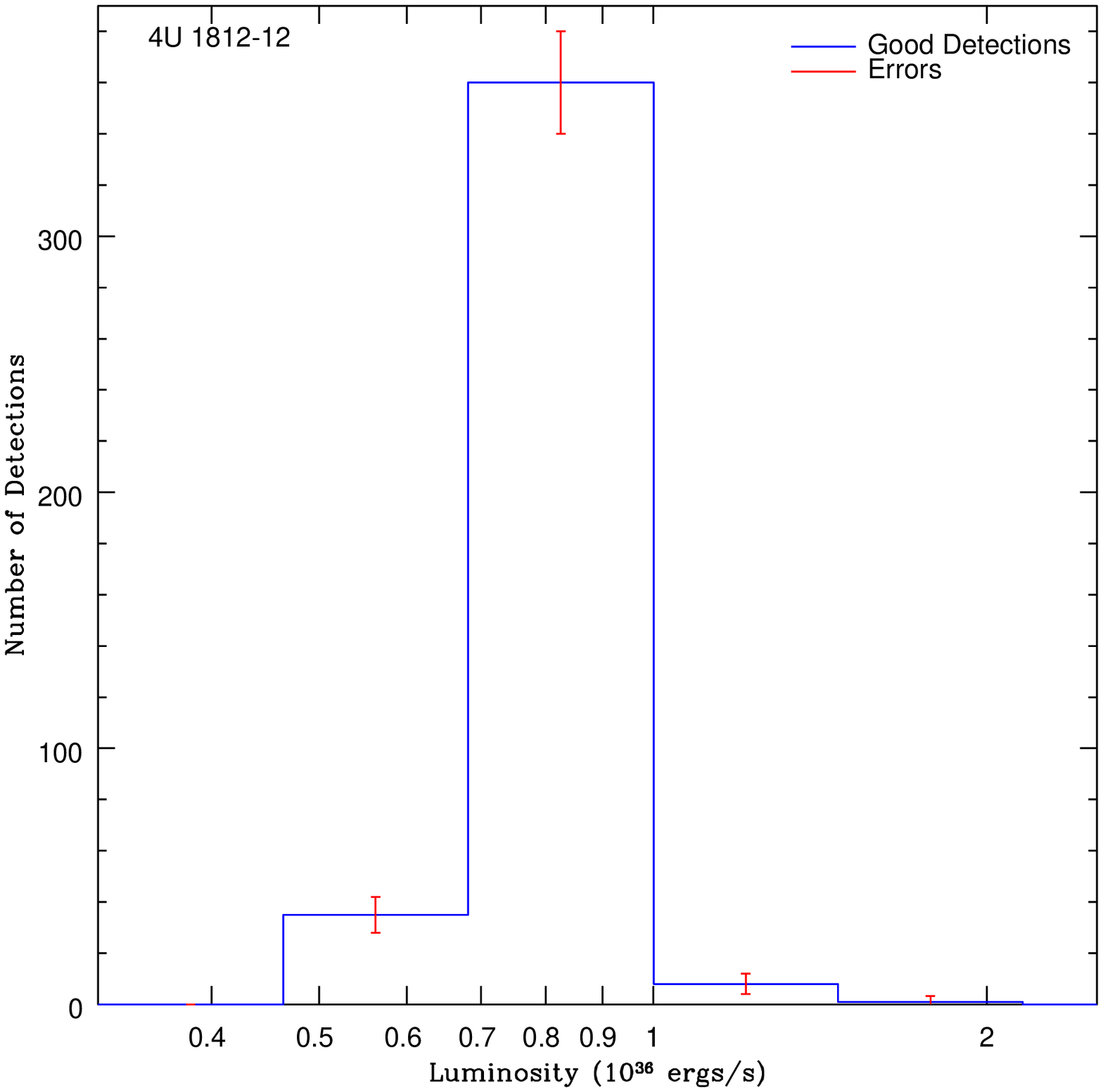} 
\end{array}{cc}$
\vspace{-4mm}$
\begin{array}{cc}
\includegraphics[scale=0.42]{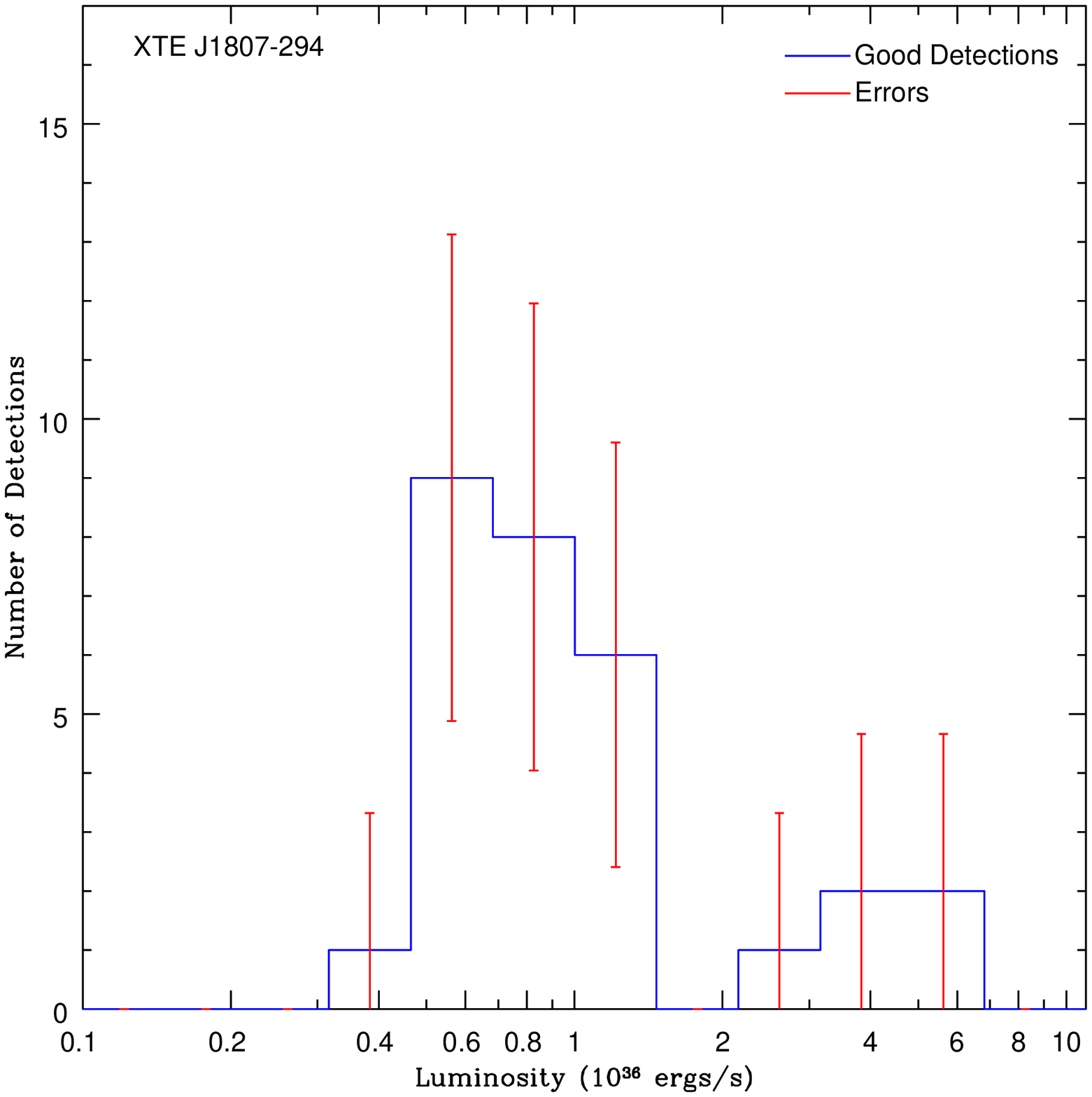}
\includegraphics[scale=0.42]{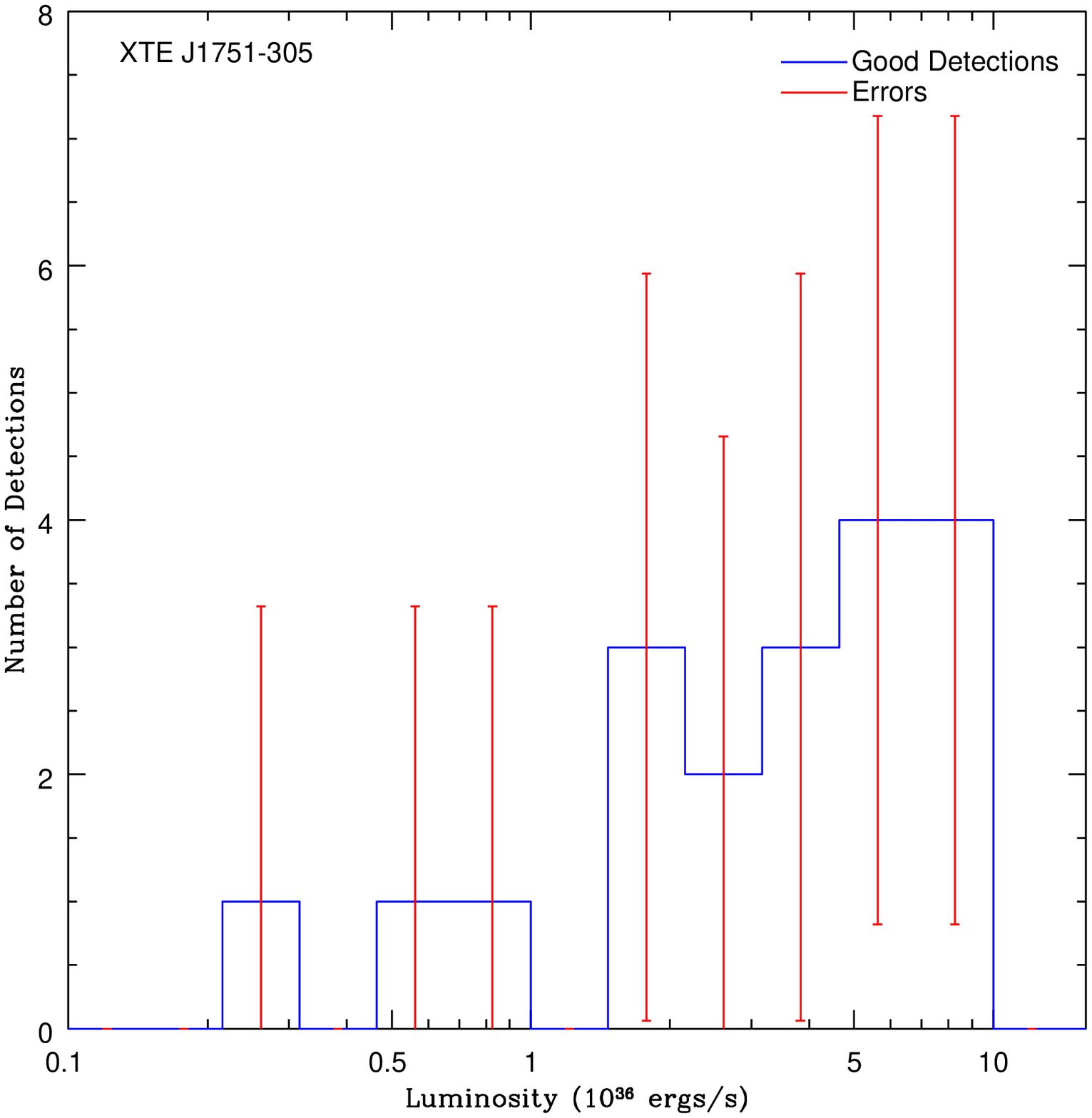} 
\end{array}{cc}$
\vspace{-4mm}
\includegraphics[scale=0.42]{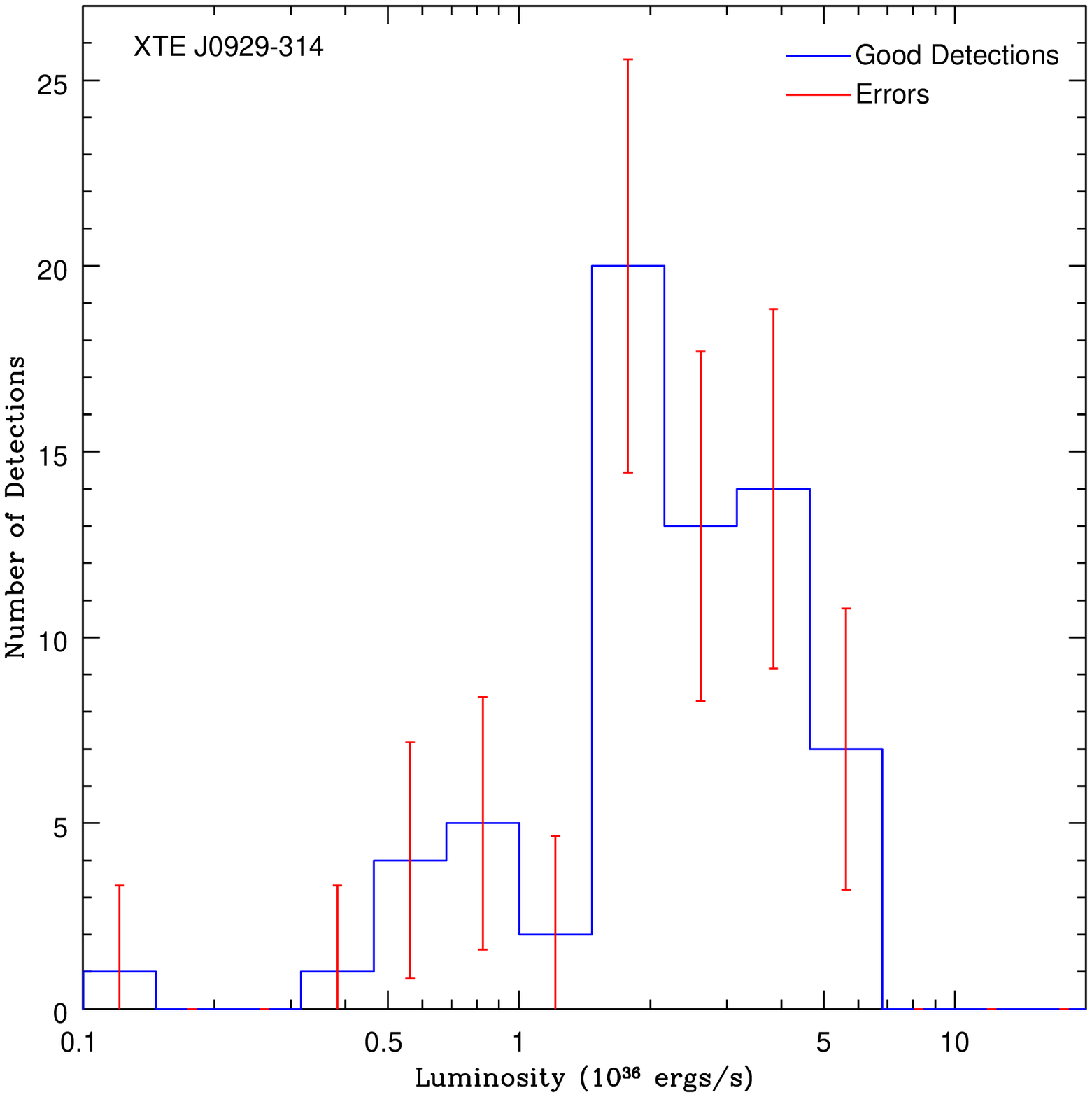} 
\addtocounter{figure}{-1}
   \caption[]{cont'd.}
\label{fig:cont3}
\end{center}
\end{figure}

\clearpage
\begin{figure}
 \begin{center}
 \figurenum{2}$
\begin{array}{cc}
\includegraphics[scale=0.42]{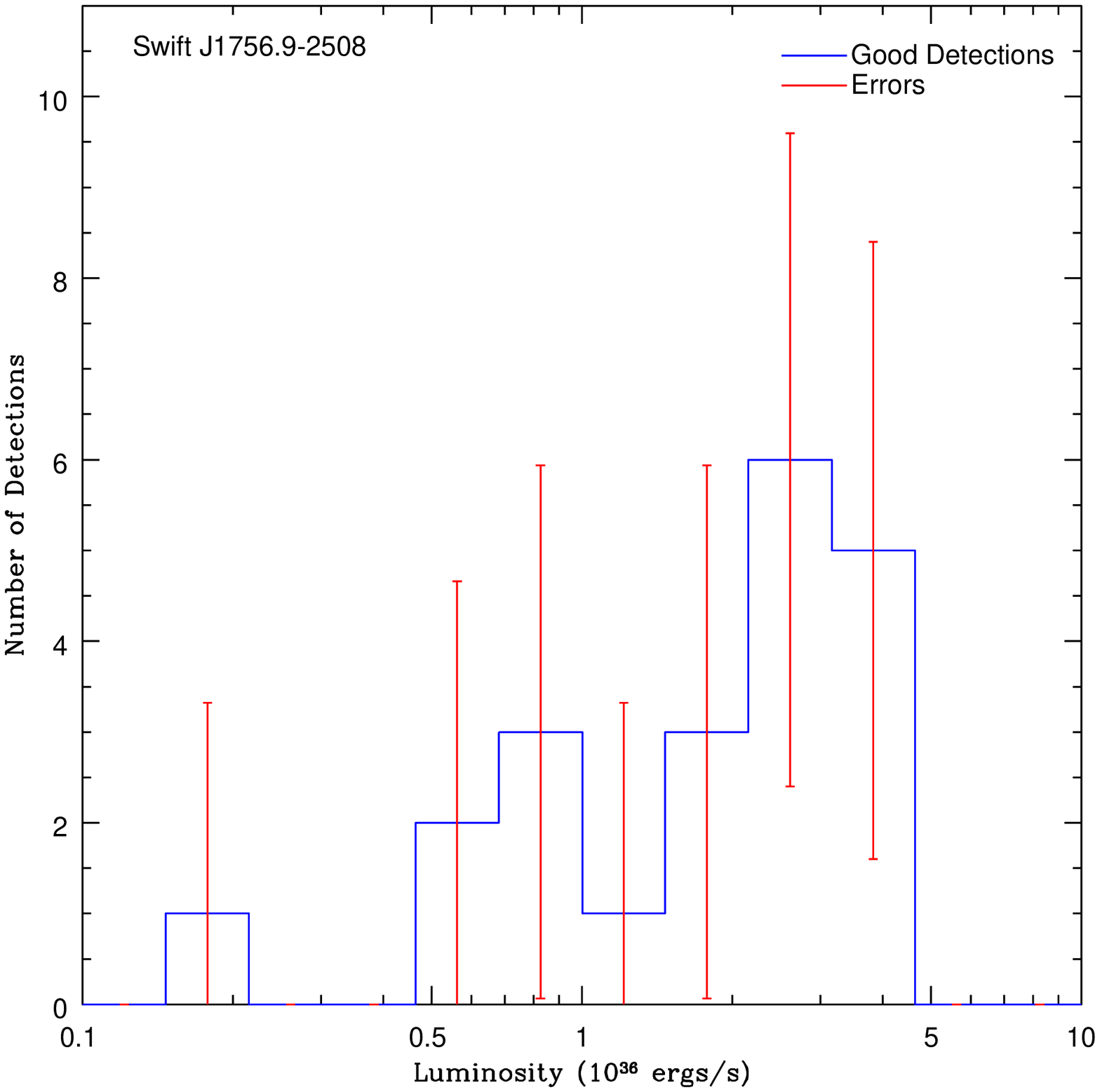} 
\includegraphics[scale=0.42]{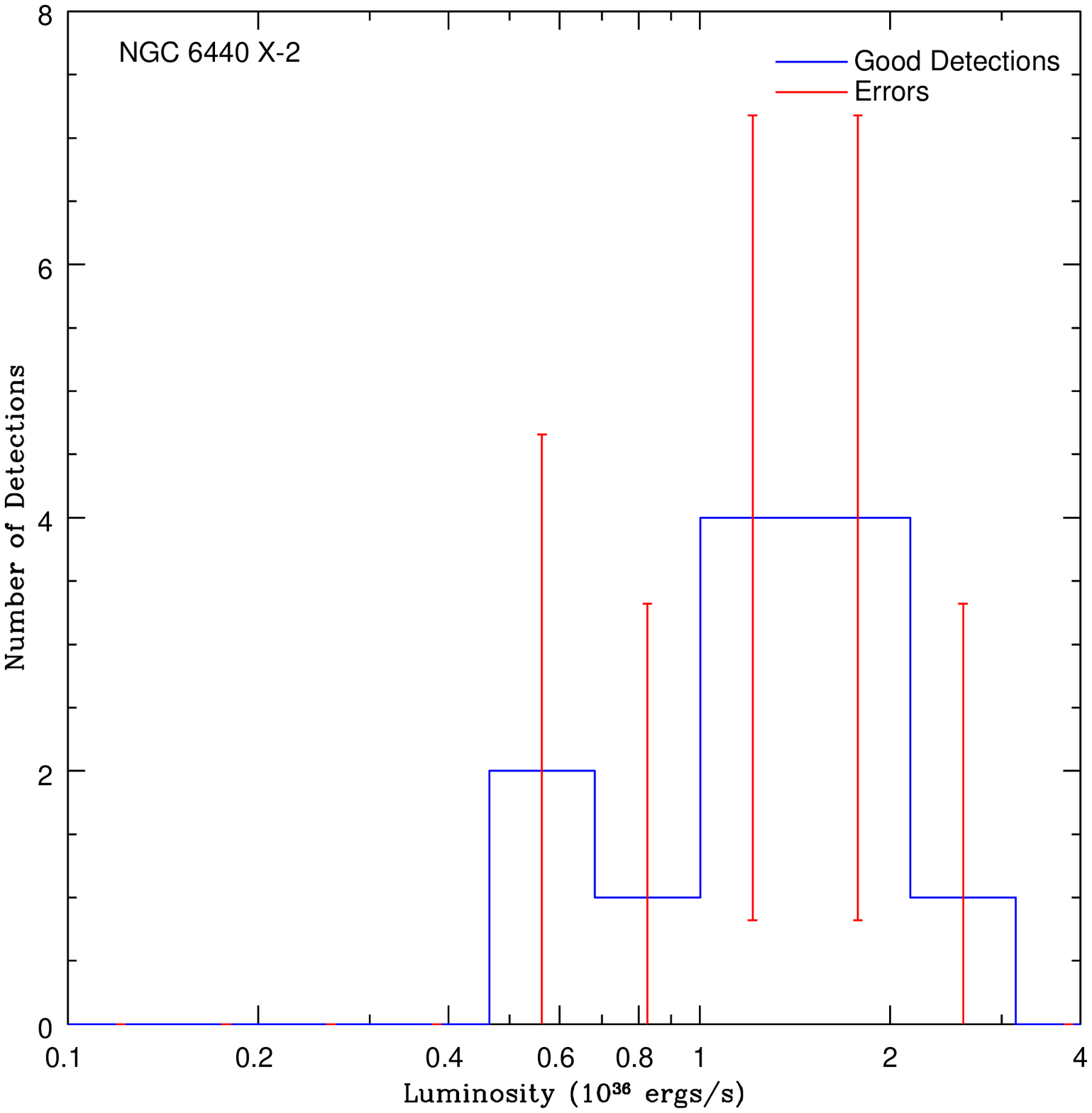}
\end{array}{cc}$
\addtocounter{figure}{-1}
   \caption{cont'd.}
\label{fig:cont4}
\end{center}
\end{figure}
  
\clearpage

\begin{figure}
\figurenum{3}
\begin{center}$
\begin{array}{cc}
\includegraphics[scale=0.4]{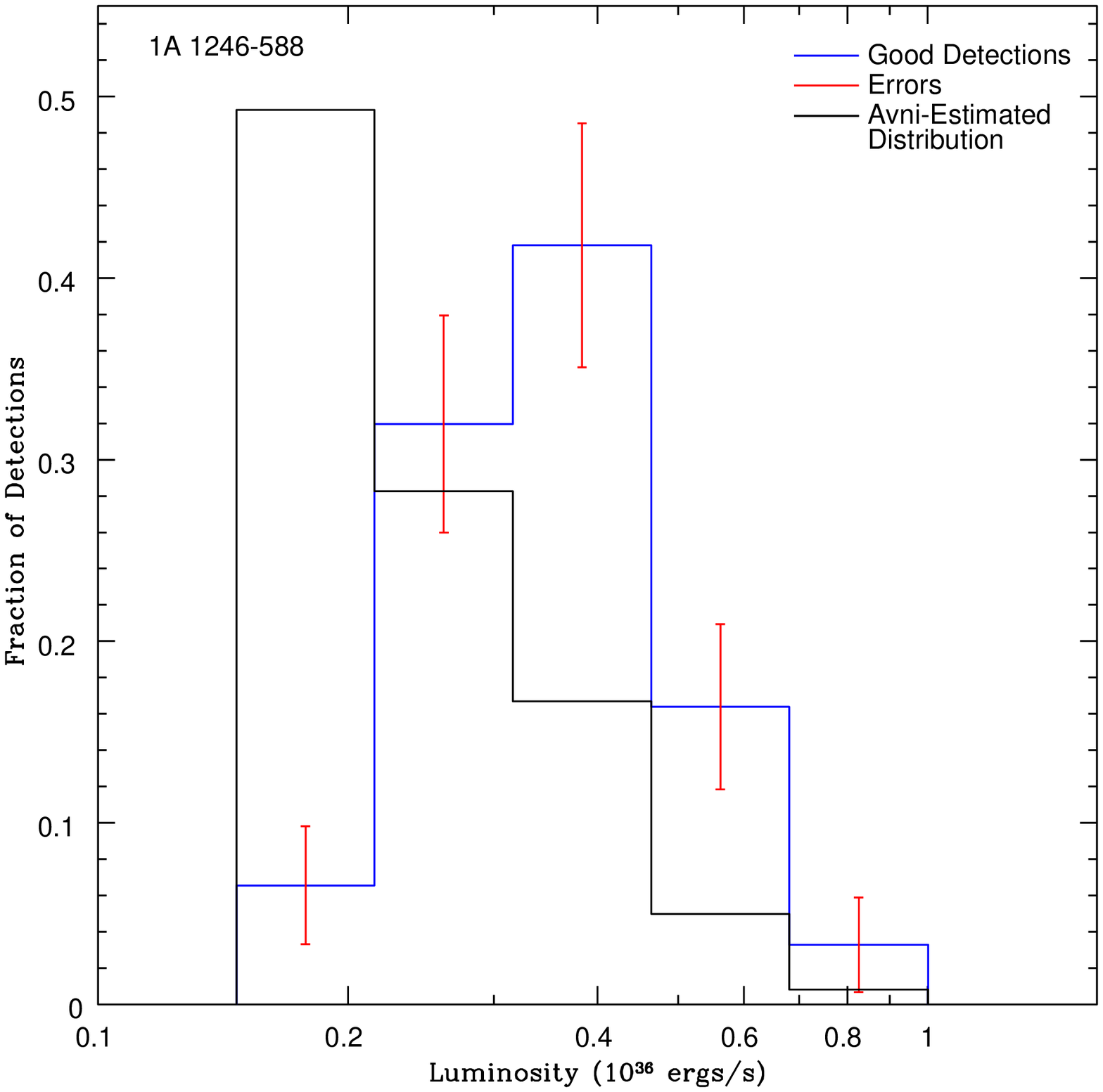} &
\includegraphics[scale=0.4]{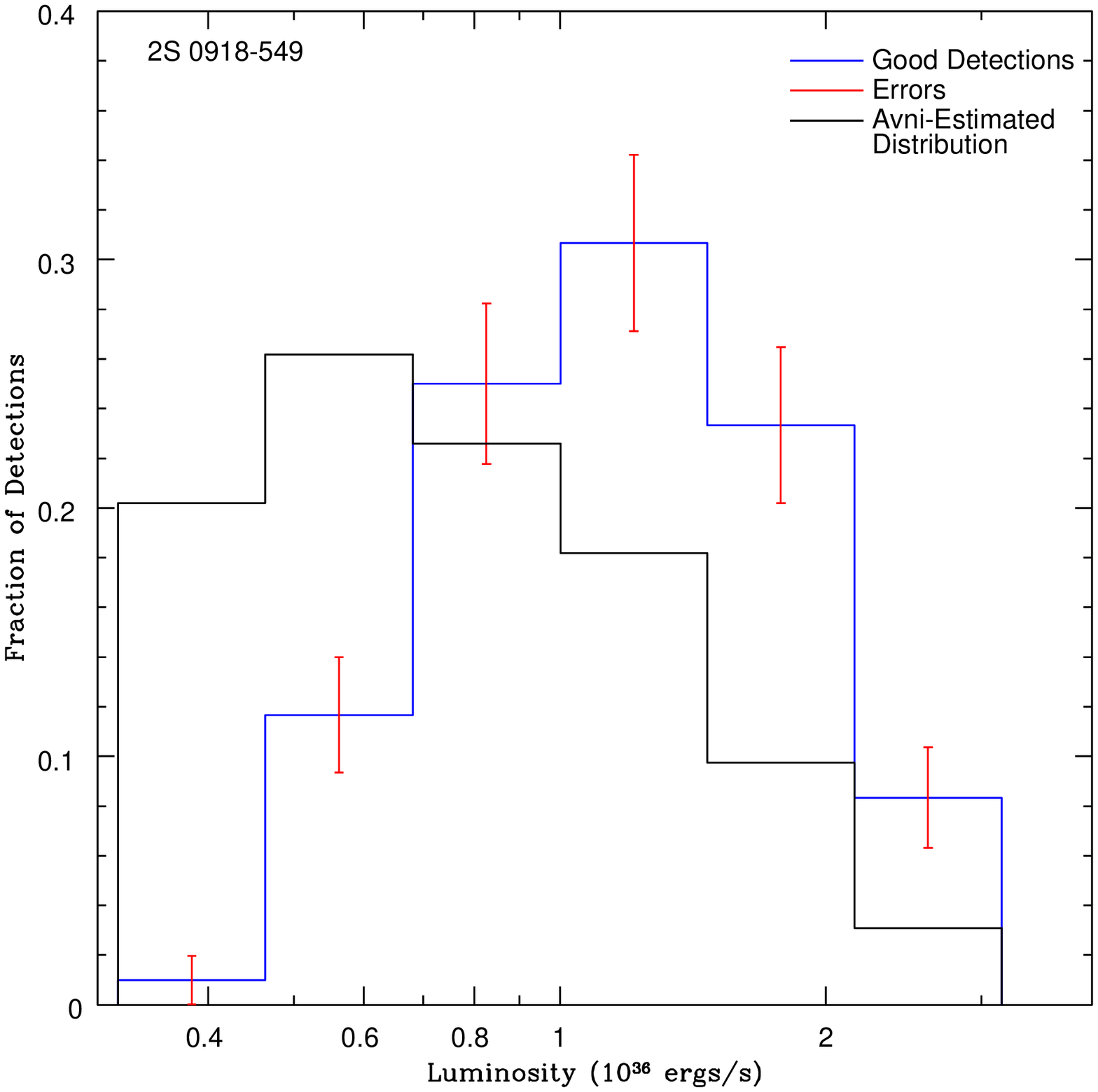} \\ 
\includegraphics[scale=0.4]{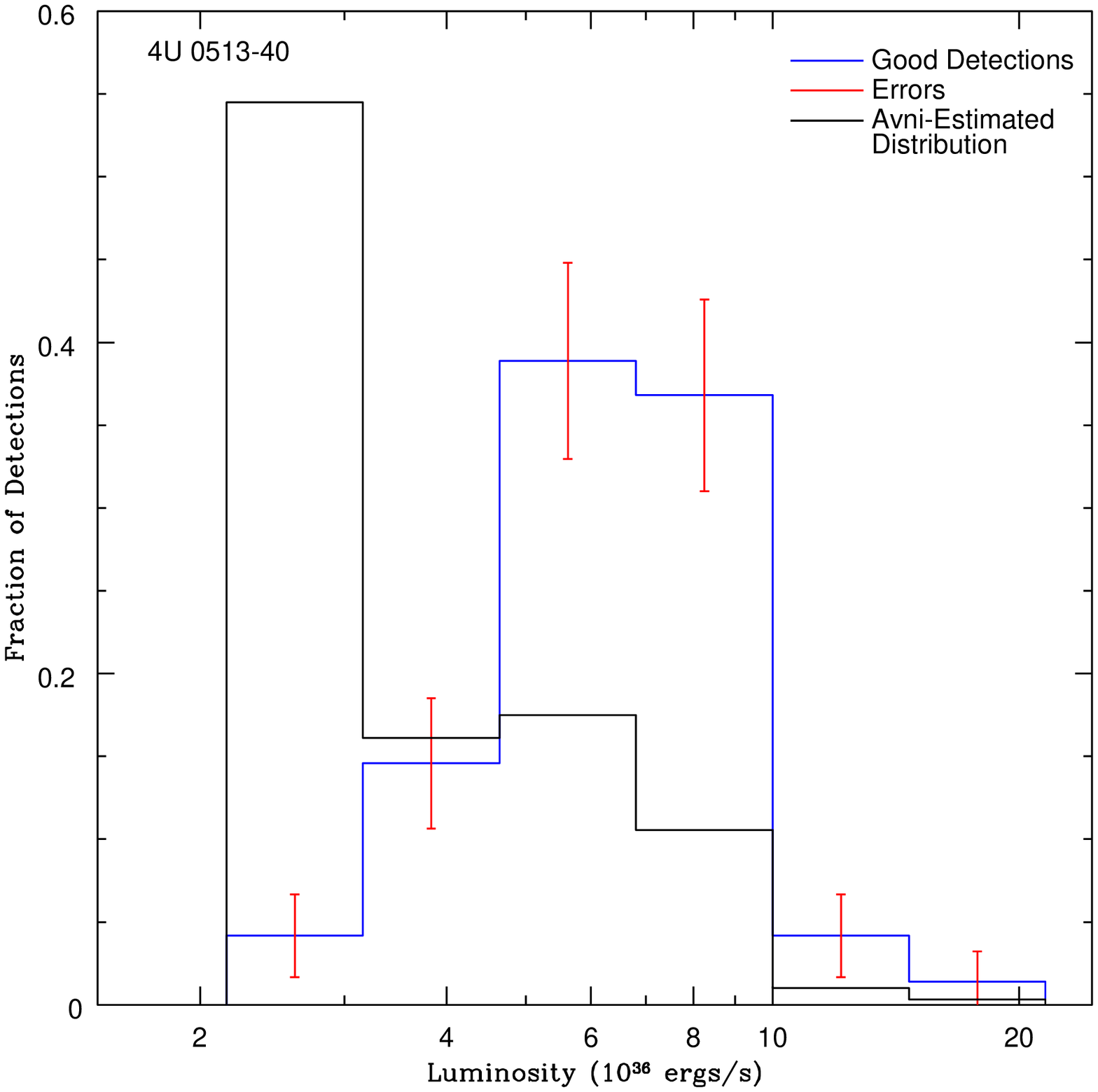} &
\includegraphics[scale=0.4]{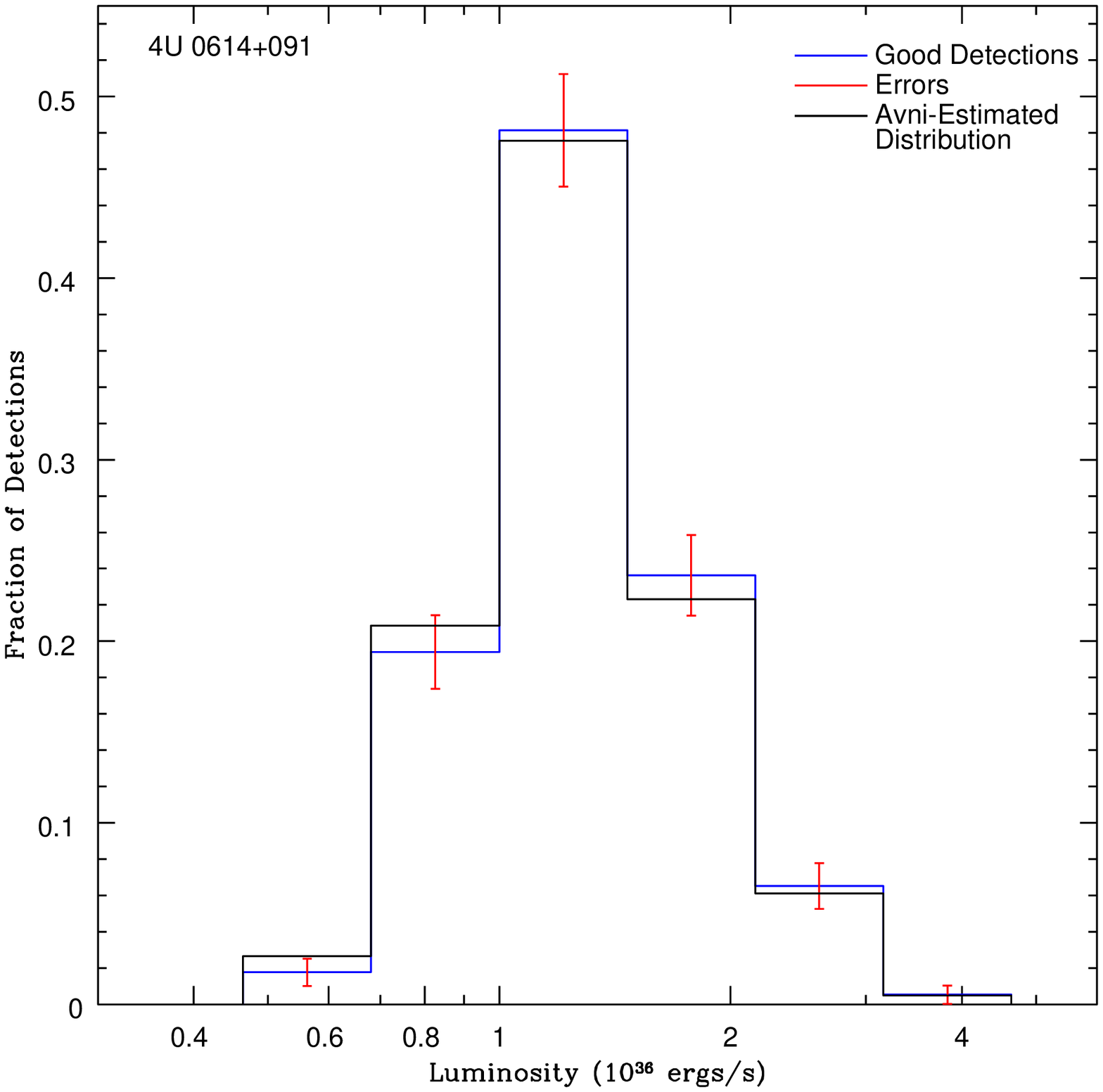}\\ 
\end{array}$
\end{center}
\label{Avni1}
\caption{Inferred X-ray luminosity functions for MAXI sources as computed from good detections and upper limits using a modified version of the maximum likelihood method of \citet{Avni80}.  The Y-axis is the fraction of all observations in that bin.
}
\end{figure}

\clearpage

\begin{figure}
\figurenum{3}
\begin{center}$
\begin{array}{cc}
\includegraphics[scale=0.4]{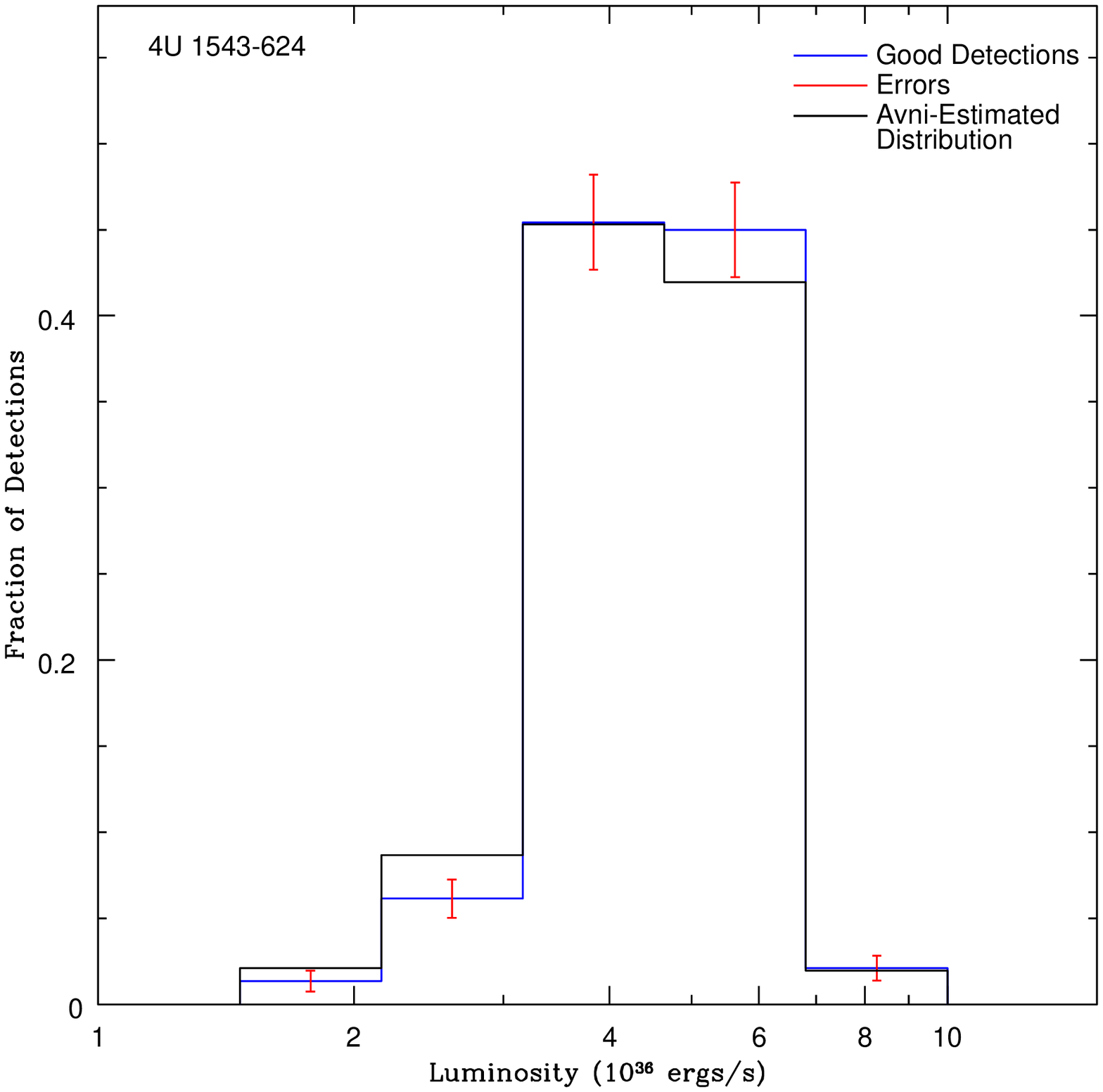} &
\includegraphics[scale=0.4]{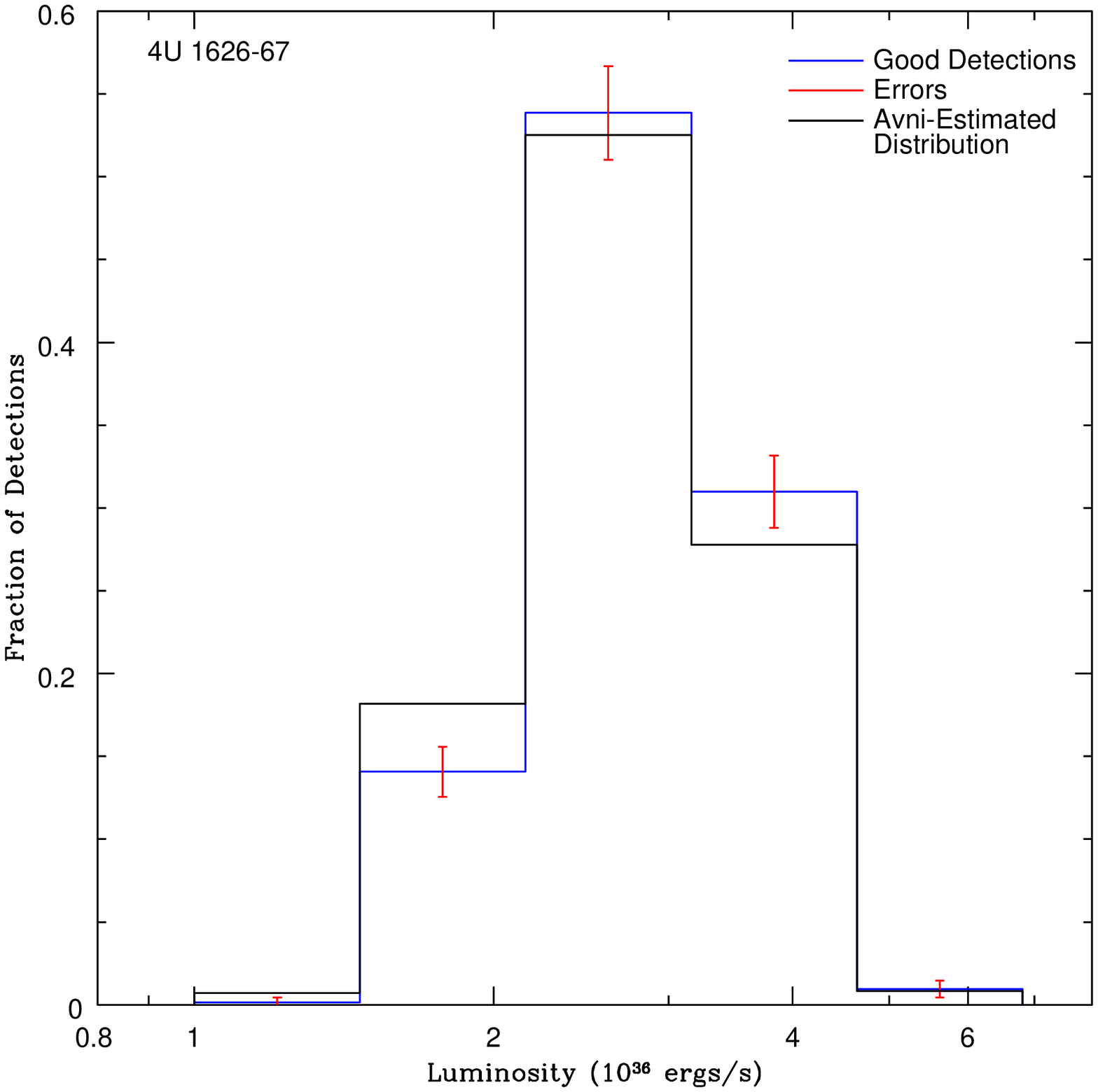}\\ 
\includegraphics[scale=0.4]{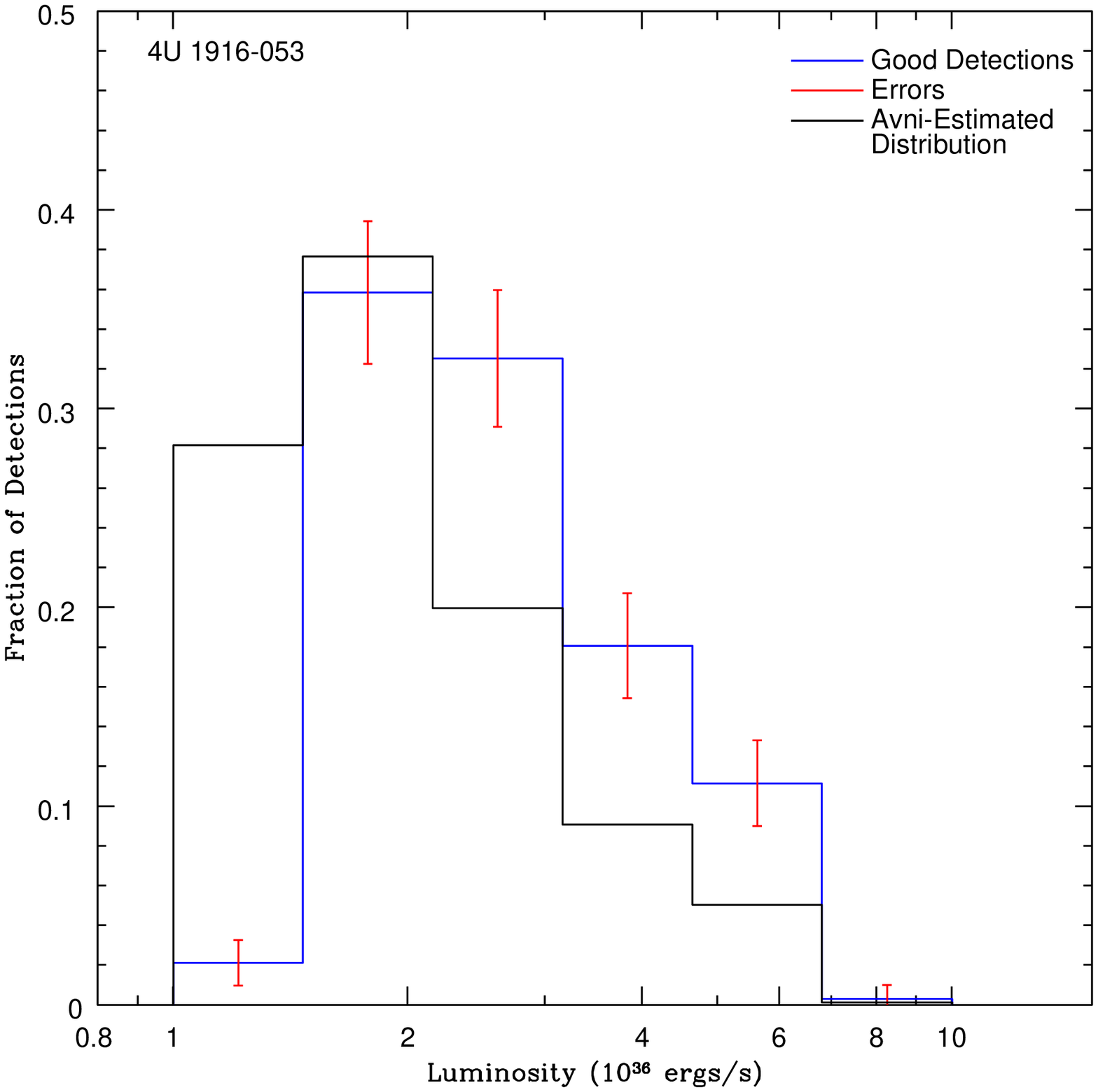} &
  \\
\end{array}$
\end{center}
\label{Avni2}
\addtocounter{figure}{-1}
\caption{cont'd}
\end{figure}

\clearpage

  \begin{table*}[h!]
\begin{centering}
\begin{tabular}{ | c@{} |@{ \hspace{0.1mm}} c |@{\hspace{0.1mm} } c@{  }|@{ \hspace{0.1mm} } c@{  }|@{\hspace{0.1mm}  } c@{  }|@{ \hspace{0.1mm} } c@{  }|}
 \hline  %
Bin Max & \multicolumn{5}{|c|}{Fraction of Detections by Source} \\
\cline{2-6}
(ergs/s) & 4U 1728-34 & 4U 1812-12 & 4U 1820-303 & 4U 1850-087 & M15 X-2 \\
\hline %
  2.15x10$^{35}$ & - & - & - & 0.005 & - \\
 3.16x10$^{35}$& - & - & - & 0.015 & -  \\
   4.64x10$^{35}$ & 0.001 & - & - & 0.144 & -  \\
 6.81x10$^{35}$ & 0.001 & 0.086 & - & 0.395 & - \\
 1.00x10$^{36}$ & 0.001 & 0.889 & - & 0.308 & 0.931\\
  1.47x10$^{36}$& 0.001 & 0.022 & - & 0.056 & -  \\
 2.15x10$^{36}$ & 0.017 & 0.003 & - & 0.031 & - \\
  3.16x10$^{36}$& 0.047 & - & - & 0.041 & - \\
 4.64x10$^{36}$ & 0.182 & - & - & 0.005 & 0.025\\
 6.81x10$^{36}$ & 0.347 & - & - & - & 0.025 \\
 1.00x10$^{37}$ & 0.224 & - & 0.001 & - & 0.014 \\
  1.47x10$^{37}$ & 0.142 & - & 0.029 & - & 0.004 \\
    2.15x10$^{37}$& 0.032 & - & 0.034 & - & - \\
 3.16x10$^{37}$ & 0.003 & - & 0.146 & -& - \\
  4.64x10$^{37}$& - & - & 0.367 & - & - \\
   6.81x10$^{37}$ & - & -& 0.392 & - & -  \\
    1.00x10$^{38}$ & - & - & 0.033 & -  & -  \\
 \hline %
 \end{tabular}
\caption{X-ray luminosity functions for the four \pca-detected sources and M15-X2, as used in random sampling calculations. All data for these sources gave secure detections.}
\label{tb:pca_xlf}
 \end{centering}
 \end{table*}

\begin{table*}[h!]
\begin{centering}
\begin{tabular}{ | c@{} |@{ \hspace{0.1mm}} c |@{\hspace{0.1mm} } c@{  }|@{ \hspace{0.1mm} } c@{  }|@{\hspace{0.1mm}  } c@{  }|@{ \hspace{0.1mm} } c@{  }|}
 \hline  %
Bin Max & \multicolumn{5}{|c|}{Fraction of Detections by Source} \\
\cline{2-6}
(ergs/s) & NCG 6440 X-2 & Swift J1756.9-2508 & XTE J1751-305 & XTE J1807-294 & XTE J0929-314 \\
\hline %
 1.00x10$^{35}$& 0.956 & 0.969 & 0.996 & 0.964 & 0.987 \\
 1.47x10$^{35}$ & - & - & - & - & 0.000 \\
 2.15x10$^{35}$ & - & 0.001 & - & - & - \\
 3.16x10$^{35}$& 0.002 & - & 0.000 & - & -  \\
   4.64x10$^{35}$ & 0.002 & 0.001 & - & 0.001 & 0.000  \\
 6.81x10$^{35}$ & 0.026 & 0.004 & 0.000 &0.011 & 0.001 \\
 1.00x10$^{36}$ & 0.001 & 0.004 & 0.000 & 0.001 & 0.001 \\
  1.47x10$^{36}$& 0.006 & 0.001 & - &0.007 & 0.001  \\
 2.15x10$^{36}$ & 0.005 & 0.004 & 0.001 & - & 0.004 \\
  3.16x10$^{36}$& 0.001 & 0.008 & 0.001 & 0.001 & 0.003 \\
 4.64x10$^{36}$ & - & 0.007 & 0.001 & 0.003 & 0.003 \\
 6.81x10$^{36}$ & - & - & 0.001 & 0.003 & 0.001 \\
 1.00x10$^{37}$ & - & - & 0.001 & - & - \\
 \hline %
 \end{tabular}
\caption{X-ray luminosity functions for the transient sources used in random sampling calculations. All upper limits are taken to indicate quiescent observations. Data for sources except for XTE J0929-314 are from the \emph{RXTE} \pca\ instrument; data for XTE J0929-314 are from \citet{Galloway02}.}
\label{transient_xlf}
 \end{centering}
 \end{table*}

\begin{center}
\begin{table*}[h!]
\begin{centering}
\begin{tabular}{ | c@{} |@{} c |@{} c@{  }|@{} c@{  }|@{} c@{  }|@{} c@{  }|@{} c@{  }| @{}c |}
 \hline  %
Bin Max & \multicolumn{7}{|c|}{Fraction of Detections by Source} \\
\cline{2-8}
(ergs/s) & 1A 1246-588 & 2S 0918-549 & 4U 0513-40 & 4U 0614+091 & 4U 1626-67 & 4U 1543-624  & 4U 1916-053 \\
\hline %
   1.00x10$^{35}$& -& - & - & - & - & - & - \\
 1.47x10$^{35}$ &- & - & - & - & - & - & -\\
 2.15x10$^{35}$ & 0.493 & - & - & - & - & - & - \\
  3.16x10$^{35}$& 0.283 & - & - & - & - & - & -  \\
   4.64x10$^{35}$ & 0.167 & 0.202 &-& - & - & - & - \\
 6.81x10$^{35}$ & 0.050 & 0.262 & - & 0.027 & - & -& - \\
 1.00x10$^{36}$ & 0.008 & 0.226 & - & 0.209 & - &- & -\\
  1.47x10$^{36}$& - & 0.182 & - & 0.476 & 0.007 & - & 0.282  \\
 2.15x10$^{36}$ & - & 0.098 & - & 0.223 & 0.182 & 0.021 &  0.377 \\
  3.16x10$^{36}$& - & 0.031 & 0.545 & 0.061 & 0.525 & 0.087 &0.200 \\
 4.64x10$^{36}$ & - & - & 0.161 & 0.005 & 0.278  & 0.453 &0.091\\
 6.81x10$^{36}$ & - & - & 0.175 & - & 0.008 & 0.420 &0.050\\
 1.00x10$^{37}$ & -& - & 0.105 & -& - & 0.020 &0.001\\
    1.47x10$^{37}$& - & - & 0.010 & - & -  & -  & - \\
 2.15x10$^{37}$ & - & - & 0.003 & -& - & - &- \\
 \hline %
 
 \end{tabular}
\caption{Inferred X-ray luminosity functions for the seven sources for which \maxi\ observations were used in our calculations, as computed using the modified maximum likelihood method of \citet{Avni80}.}
\label{Avni_XLFs}
\end{centering}
 \end{table*}
 \end{center}
 
\begin{figure}[h!]
\figurenum{4}
\includegraphics[scale=0.45]{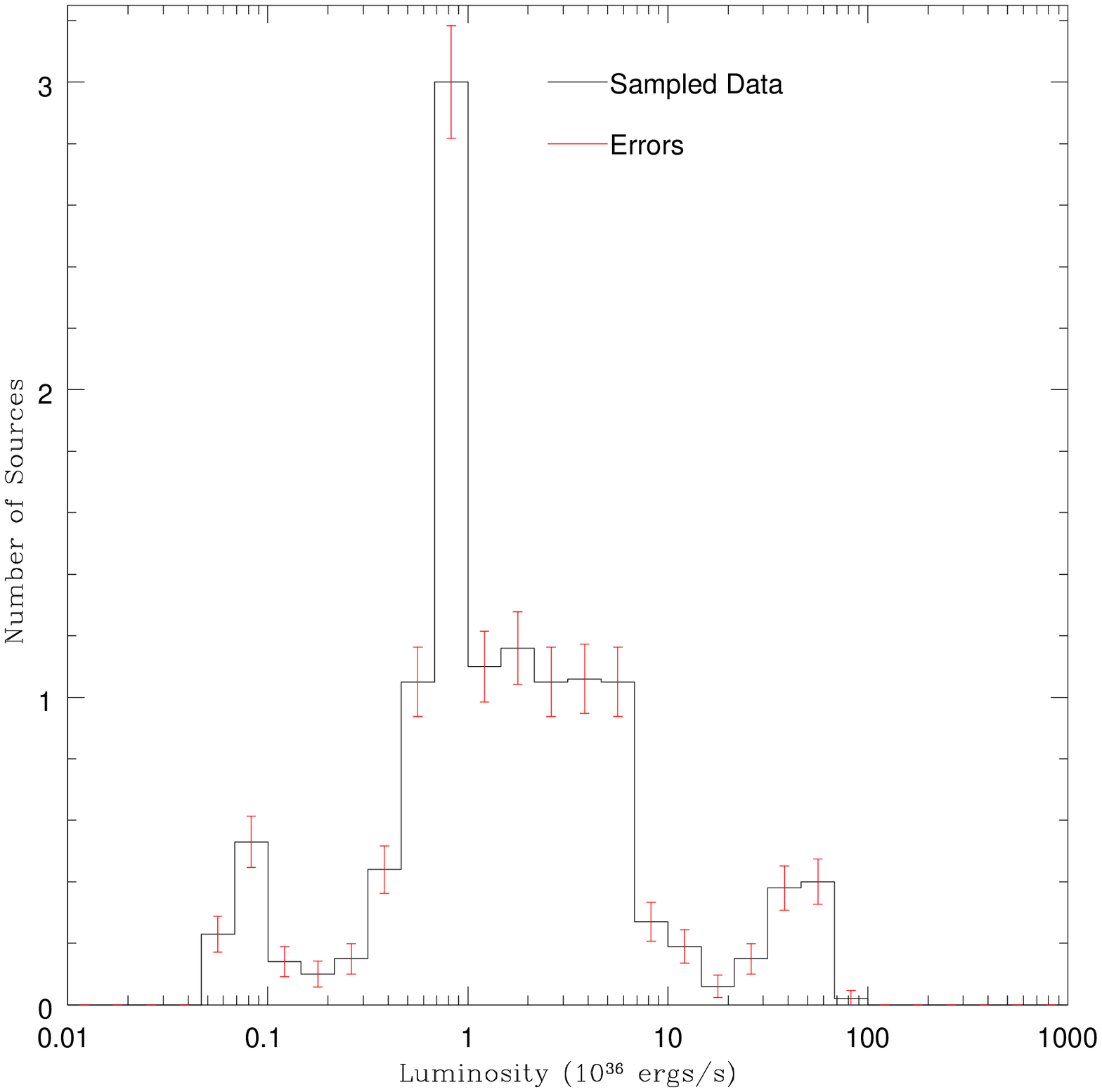}
\caption{Results of extracting 100 data points at random from each of 17 sources' XLFs to make a combined luminosity function for UCXBs. Errors from \citet{Gehrels86}. Counts in each bin were divided by 100 to produce the plot, whose ordinate is in units of sources (out of the 17 total used for sampling). 
 }
\label{xlf_total}
\end{figure}

We 
randomly selected 100 observations from each UCXB's XLF. 
The 1700 data points were also combined to create a combined luminosity function of the population.  Since the transients are rarely in outburst, they contribute little to the total UCXB luminosity function; most values drawn from their distribution are '0'. Fig.\  \ref{xlf_total} displays the combined luminosity function, with \citet{Gehrels86} errors plotted.
 The y-axis is in units of sources--including the quiescent points (not shown), the sum of all bins comes to 17.  
Repeated samples yielded similar histograms.  Samples computed using only good detections (rather than the Avni-computed XLFs) for the \maxi\ sources showed only minor differences in bins greater than the peak, and the peak location did not change.

We fitted the slope of the 100 individual samplings of the XLF, and of the combined XLF, with a simple power-law in Sherpa, using the C-statistic due to the low (or zero) numbers of counts in many bins.  We computed Gehrels errors on the number of counts in each bin, and then put the XLF into a differential, dN/dL format for fitting, by dividing the number of counts per bin by the bin width.  
We fit  $7\times10^{35}$$<L_X<$$10^{38}$ ergs/s.  
 Less luminous bins are significantly incomplete, due to the limitations of all-sky monitors.  (We suffer incompleteness in identification of UCXBs in all bins, but that incompleteness is not obviously luminosity-dependent, and is difficult to quantify.)  The 100 individual samplings produced acceptable power-law fits, with indices of $\alpha$ (for $dN(L)/dL=k L^{-\alpha}$) ranging from 0.92 to 2.57, 
with a mean of 1.66 and error of 0.28 (Fig. 5).
The combined XLF was poorly fit with a power-law (Fig. 6; index 1.65, probability 8e-26, reduced statistic 9.6), likely due to the limited number of sources producing a ``lumpy'' XLF.  However, we did obtain the same power-law index as the mean of the individual samples (1.65).  The ``lump'' above $10^{37}$ ergs/s, for instance, is produced by 4U 1820-30.  Thus a more complex fit would not be meaningful.

A relevant question is whether the XLF of globular cluster UCXBs differs from that of field UCXBs. 
Unfortunately, our statistics are too small to significantly test this question, as we have only five globular cluster UCXBs (one of them transient).
Although the globular cluster UCXBs include the highest luminosity system (4U 1820-30), the other globular UCXBs are not unusual compared to the field systems, or to other suspected ultracompact globular cluster systems (e.g. in Terzan 2, \citealt{intZand07}).

\begin{figure}
\figurenum{5}
\includegraphics[scale=0.43]{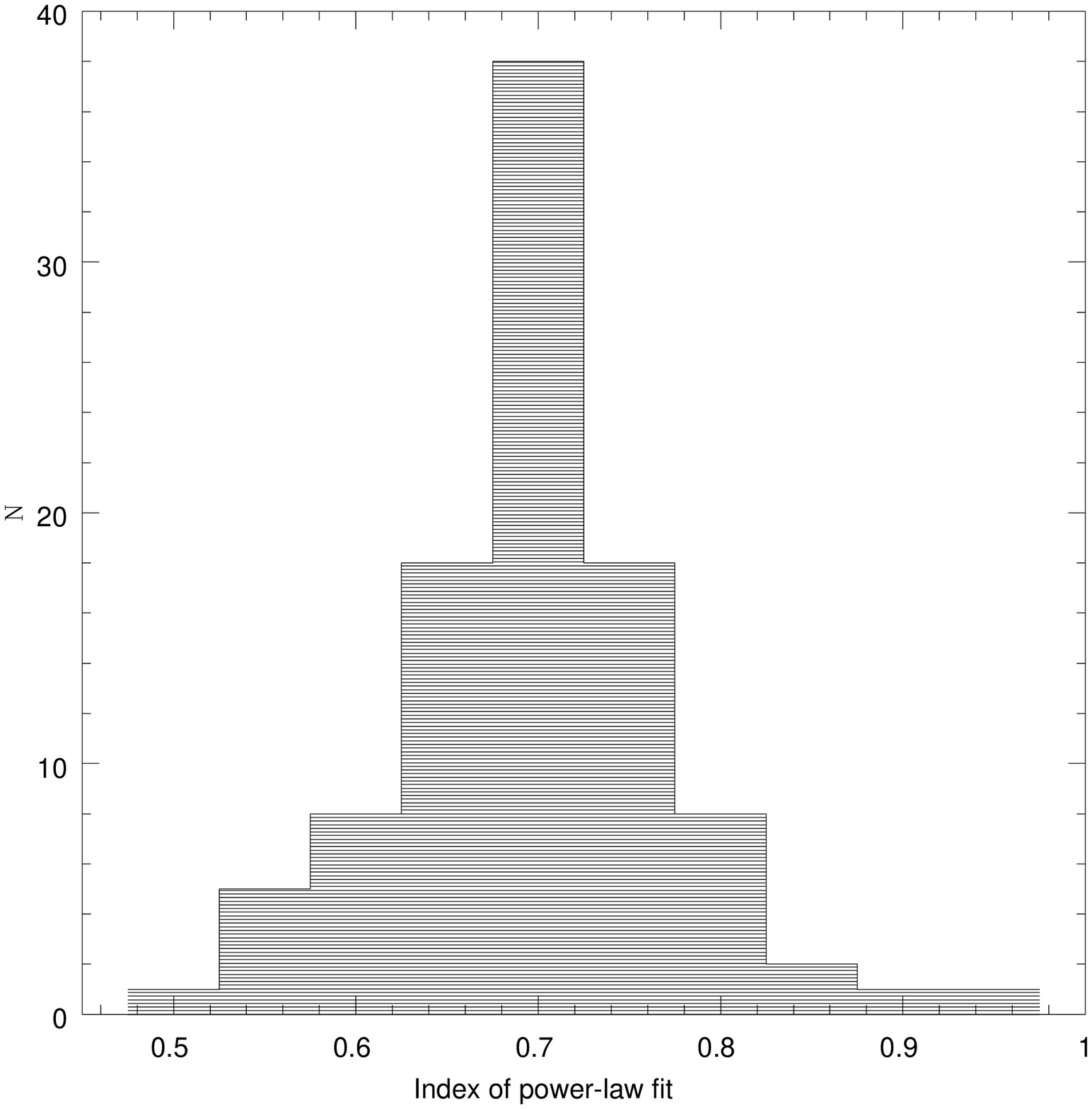}
\caption{Histogram of power-law indices $\alpha$ fit to 100 differential XLF samplings ($dN(L)/dL=k L^{-\alpha}$).  Typical index errors per fit are about 0.3. 
}
\label{index_hist}
\end{figure}

\begin{figure}
\figurenum{6}
\includegraphics[scale=0.75]{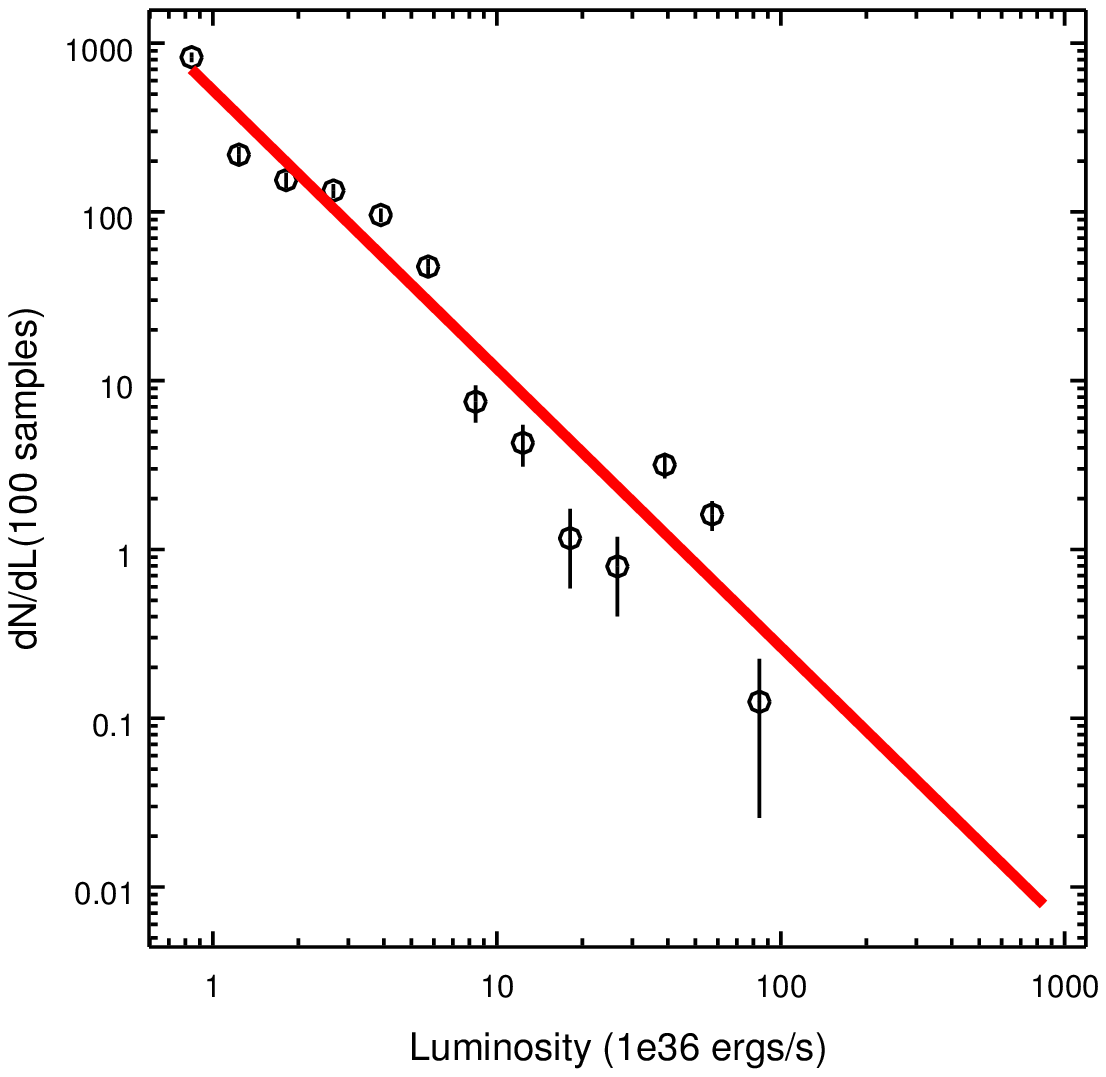}
\caption{  
The combined `population' differential XLF, compiled using 100 randomly-drawn data points from each of the 17 sources' XLFs. dN/dL was computed for each bin by dividing the effective number of sources per bin by the bin width.  A power-law of index $\alpha$=1.65 ($dN(L)/dL=k L^{-\alpha}$) fits the general slope, although the XLF has lumps due to individual sources.
}
\label{combined_sherpa}
\end{figure}

\subsection{Comparing to RXTE ASM studies of galactic UCXBs}
We note that the recent work of \citet{vanHaaften12} on the luminosity functions for galactic UCXBs, using individual \RXTE\ ASM measurements, argues for rapid variations of all sources and finds luminosity estimates that differ from ours.  Their analysis is an important contribution, but we have two concerns about the analysis procedures.  First, systematic errors in the ASM data are not considered, which may affect both their individual dwell measurements (by increasing the number of spurious datapoints) and the countrates averaged over all dwells.  Two of their sources show negative countrates averaged over all dwells, evidence for the existence of systematic errors.  

Our other concern is that luminosities approaching $10^{38}$ ergs/s for $10^{-4}$ of the observations suggest thermonuclear X-ray bursts, including normal short helium-burning bursts or, for lower mass accretion rates, intermediate-duration (up to half-hour) X-ray bursts \citep{Galloway08,intZand05}.  We can roughly estimate the effect on 90 second dwells of short bursts if we know the recurrence time and $\alpha$ (the ratio of persistent to burst fluence).  4U 1728-34, for instance, shows short bursts with decay timescales of 6.3~s, recurrence times of 2.5-5~hours, and $\alpha$ averaging 150 (from 90--300).  Thus, each burst will affect one ASM datapoint, they will affect $5\times10^{-3}$--$10^{-2}$ of the dwells, and they will increase the flux by typically a factor of 2.  Fainter systems will typically show rarer bursts, with larger fractional increases.  Intermediate-duration bursts and superbursts last longer than an ASM dwell, reach the Eddington limit, and are much rarer (e.g. intermediate-duration bursts appear once per 85~days for SLX 1737-282, \citealt{Falanga08}).  Bursts from UCXBs show a range of fluences between these extremes, which may substantially affect the observations of \citet{vanHaaften12}. Clearly more detailed analysis of the unique RXTE/ASM dataset is suggested, and may provide deep insight into UCXB burst properties.

\section{Discussion}

\subsection{Prevalence of strong variability in persistent UCXBs} 
The origin of substantial (up to a factor of 10) variability seen in a number of persistent UCXB sources is not understood. \citet{Maccarone10} highlighted such variations in 4U 0513-40.  Similar nonperiodic variations of up to a factor of 10 have been identified in several persistent UCXBs:  
4U 1915-05 \citep{Simon05}, 1A 1246-58 \citep{intZand08}, 4U 1626-67 \citep{CameroA10}, M15 X-2 \citep{Sivakoff11}, \& \citet{intZand07}'s (strong candidate) UCXBs SAX J1712.6-3739, SLX 1735-269, \& 1RXS J170854.4-321857 \citep{intZand05}.  We now add 4U 1850-087, 2S 0918-549, and 4U 1728-34 to the list. 
 Slightly smaller flaring (of up to a factor of 5) is seen during times when the persistent flux is at  particular values for  
4U 0614+09 \citep{Kuulkers10} and 
\citet{intZand07}'s strong UCXB candidate 4U 1722-30.
\citet{intZand07} discusses an explanation of the rapid flaring behavior as a tidal instability due to a precessing accretion disk \citep{Whitehurst88}, with a possible dependence on mass transfer rate to explain its presence/absence in different epochs for, e.g., 4U 1722-30.  
Such variations may not exist for high-luminosity ($L_X>3\times10^{37}$ ergs/s) ultracompacts, with their presumably more massive donors in shorter orbits \citep{Bildsten04}, though 4U 1820-30's well-known cycles illustrate the existence of strong variability in some such systems.  The prevalence of strong variation reinforces \citet{Maccarone10}'s point about the importance of considering variation among persistent systems when identifying extragalactic transients, and thus the importance of empirical observations of the luminosities of systems of known type for validating luminosity function modeling.

\subsection{Luminosity functions of UCXBs}
We have produced an empirical X-ray luminosity function from known UCXBs in the Milky Way.  The list of Galactic UCXBs is certainly incomplete and may suffer biases (for instance, we are probably missing persistent and transient sources below $10^{36}$ ergs/s, so we advise readers to treat our luminosity function for bins below this simply as upper limits). However, it is important to test theoretical calculations against the best data available. We have few sources, and typically less than one above $L_X=10^{37}$ ergs/s (see Fig. \ref{xlf_total}), which makes comparison of this UCXB luminosity function with those of LMXBs in distant galaxies difficult.   However, a few nearby galaxies have been studied sufficiently to make a useful comparison.

First, we note that the slope of our X-ray luminosity function (1.66$\pm$0.28 for the samplings, 1.65 for the cumulative) down to $\sim10^{36}$ ergs/s is in good agreement with the predicted slope for ultracompact X-ray binaries (1.7--1.8, depending on the white dwarf entropy) found by \citet{Bildsten04}.   This is not a trivial result, since the theoretical calculation did not include strong variability among persistent systems, transients, or a mixture of UCXBs with different formation mechanisms and disk instability lines (needed to explain, for instance, the $P_{\rm orb}$=40--60 minute high-mass-transfer systems; see Paper II).  

Our result does not agree with simulations of the UCXB XLF presented in \citet{Fragos08}, which have similar slopes but cut off at 5$\times 10^{36}$ ergs/s for both persistent and transient UCXB systems.  Our results indicate that the UCXB XLF continues down to at least 1$\times10^{36}$ ergs/s.  The cutoff in \citet{Fragos08} is physically motivated by the criterion for disk instability of irradiated pure helium disks, calculated in \citet{Lasota08}, and assumed to be appropriate for a population of UCXBs.  Here, we show empirical evidence against this cutoff; in Paper II we identify the physical reasons behind the lowered disk instability limit.  

The agreement of the theoretical UCXB slope with the XLF slope of several known elliptical galaxies suggested that ultracompact binaries dominate the XLF in globular clusters \citep{Bildsten04}.  However, deeper observations have shown that the XLFs of old populations break to shallower slopes at luminosities below $10^{37}$ ergs/s, and that globular clusters have a significantly shallower slope at low luminosities.  These results have been shown for the bulge and globular cluster system of M31 
\citep{Kong03,Trudolyubov04,Voss07}, Cen A \citep{Voss09}, NGC 3379 \citep{Fabbiano07}, and for combined studies of several nearby galaxies \citep{Zhang11}.
 In particular, the similarity of the slope and $10^{37}$ ergs/s cutoff of the observed Cen A globular cluster XLF to the theoretical UCXB XLF (cut off at $5\times10^{36}$ ergs/s in \citealt{Fragos08}) motivated \citet{Voss09} to suggest that the Cen A globular clusters were dominated by UCXBs.  Although this was an excellent idea (given the enhanced population of UCXBs in Milky Way globular clusters), our UCXB XLF indicates that this is not feasible.  Other suggested explanations of the difference are significantly more complicated \citep{Fragos08,Kim09,Zhang11}.  

The steep XLF we find down to $L_X=10^{36}$ ergs/s, combined with the theoretical expectation that this XLF will continue to higher $L_X$ and the relatively flat XLFs of both bulge and globular cluster XLFs between $10^{36}<L_X<10^{37}$ ergs/s, indicate that UCXBs make up only a minor portion of bulge and globular cluster XLFs above a few $\times10^{36}$ ergs/s.

\section{Conclusions}

We have constructed luminosity functions for clearly identified UCXBs from the best available long-term data, using, in order of priority, \RXTE\ \pca\ bulge scans, \maxi\ lightcurves, \RXTE\ ASM lightcurves supplemented with \Chandra\ observations, and literature reports of the luminosities of systems during transient outbursts. Variability by up to a factor of 10 is quite common among persistent UCXBs, which is important to consider when identifying transient behavior in other galaxies \citep{Maccarone10}.
 We have taken account of upper limits in constructing these luminosity functions, using the method of \citet{Avni80} to calculate a best estimate of the true luminosity function.  We combined this information to make ``snapshot'' luminosity functions (one sample from each source) and a combined luminosity function using 100 samples from each source.  Both methods find consistency with a power-law of slope $\alpha=1.66\pm0.28$, extending down to $10^{36}$ ergs/s (below which we are incomplete).  

Our empirical UCXB luminosity function extends down in luminosity without a break to $\sim10^{36}$ ergs/s, which disagrees with current theoretical luminosity functions that cut off at $5\times10^{36}$ ergs/s. The slope, however, is in agreement with theoretical predictions from UCXB evolution.  Comparing our UCXB luminosity function to current measurements of luminosity functions in old stellar populations and globular clusters in other galaxies suggests that UCXBs make up only a small fraction of the X-ray binary population in either case.

\acknowledgments

This research has made use of \maxi\ data provided by RIKEN, JAXA and the \maxi\ team, \pca\ bulge scan data provided by C. Markwardt and the \RXTE\ team, \RXTE\ ASM data products provided by the \RXTE\ team, \Chandra\ archival data, and the ADS. We acknowledge financial support from NSERC (Discovery Grants to COH, NI, and GRS, and an NSERC USRA, Julie Payette NSERC Research Scholarship, and Andr\'e Hamer Postgraduate Prize supporting MCE), an Alberta Ingenuity New Faculty award to COH, a Canada Research Chair supporting NI, and the Avadh Bhatia Fellowship supporting JCG.

\bibliography{src_ref_list}
\bibliographystyle{apj}

\section{Appendix A}

Consider $M$ luminosity bins indexed by $n$. If our data consisted only of valid detections, the value of the distribution function for the $n^{th}$ bin, $f_n$, would be straightforwardly given by 

\begin{equation}
f_n = \frac{N(n)}{J} 
\end{equation}

where $N(n)$ is the number of valid detections in the $n^{th}$ bin, and $J$ is the total number of detections in all $M$ bins. Since we are dealing with upper limits as well as valid detections, we instead need to define for each of the $n$ bins an ``effective'' number of detections, which combines the valid detections and the most likely distribution of the upper limits, 

\begin{equation}
f_n = \frac{N_{\rm eff}(n)}{J} 
\label{effective_dist}
\end{equation}

where $N_{\rm eff}(n)$ is an effective number of detections in the $n^{th}$ bin. To calculate this effective number, the nature of the most likely distribution of upper limits needs to be considered.

An upper limit detection with luminosity value $L$ that falls into the $k^{th}$ bin could be random noise or a true detection with a luminosity value anywhere from 0 up to the value $L$. Thus, the \textit{true} value could fall in any of $n$ bins, with $n \le k$. If we consider an individual such bin $n$, it can contain a portion of the upper limits in \textit{all bins with $k \ge n$}. To find the total number of upper limit counts that truly correspond to bin $n$, $T(n)$, then, we must perform the following sum:

\begin{equation}
T(n)= \sum_{k=n}^{M}\frac{U(k) f_n}{\sum_{z=1}^{k} f_z}
\end{equation}

Where $U(k)$ is the number of upper limit detections which fall into the $k^{th}$ bin. In the maximum likelihood scenario, the way that these $U(k)$ upper limit counts are distributed amongst all lower bins depends on the distribution function - that is, the probability bin $n$ would contain a count divided by the total probability for all bins lower than $k$. We must take the outer sum, over all bins with $k \ge n$, because upper limit counts from all of these bins contribute to ``true" counts in bin $n$. 

Once we know the way in which upper limits are likely to be distributed among the bins, it is a simple matter to derive $N_{\rm eff}(n)$:

\begin{equation}
N_{\rm eff}(n) = N(n) +\sum_{k=n}^{M}\frac{U(k) f_n}{\sum_{z=1}^{k} f_z}
\label{Neff}
\end{equation}

Where we have simply added the valid detections to the most likely number of upper limits that actually lie in bin $n$. 

Now, returning to equation \ref{effective_dist} and substituting equation \ref{Neff} for $N_{\rm eff}(n)$, we obtain:

\begin{equation}
f_n = \frac{N(n) +\sum_{k=n}^{M}\frac{U(k) f_n}{\sum_{z=1}^{k} f_z}}{J}
\label{fnalmost}
\end{equation}

Isolating $f_n$,

\begin{equation}
f_n = \frac{N(n)}{J-\sum_{k=n}^{M}\frac{U(k)}{\sum_{z=1}^{k}f_z}}
\end{equation}

Finally, we note that we can rewrite the sum in the denominator as follows, using the fact that by definition, all the probabilities in the distribution sum to 1:

\begin{equation} 
\sum_{z=1}^{k}f_z = \sum_{z=1}^{M}f_z - \sum_{z=k+1}^{M}f_z = 1 - \sum_{z=k+1}^{M}f_z
\end{equation}

Making the above replacement, we thus obtain

\begin{equation}
f_n = \frac{N(n)}{J-\sum_{k=n}^{M}\frac{U(k)}{1-\sum_{z=k+1}^{M}f_z}}
\end{equation}

This formula is clearly recursive, due to the presence of the distribution function values on the right hand side. If we start with bin $M$,  the sum in the denominator disappears (which makes sense, as there are no higher bins to potentially contribute their upper limits). Once we have $f_M$, we can use it to calculate $f_{M-1}$ and so on, proceeding from the highest to the lowest bin recursively. 

\end{document}